\documentclass[aps,prb,numerical,reprint,superscriptaddress,showpacs,floatfix,longbibliography]{revtex4-1}
\usepackage{graphicx}
\usepackage{dcolumn}
\usepackage{bm}
\usepackage{color}
\usepackage{epstopdf}
\usepackage{multirow}
\usepackage{soul}

\begin{document}

\preprint{PRB/GaN(1-100)+H}

\title{Adsorption and desorption of hydrogen at nonpolar GaN$(1\overline{1}00)$ surfaces: Kinetics and impact on surface vibrational and electronic properties}

\author{L. Lymperakis}
\email{lymperakis@mpie.de}
\affiliation{
Max-Planck-Institut f\"{u}r Eisenforschung GmbH, Max-Planck-Stra{\ss}e 1, 40237 D\"{u}sseldorf, Germany}%

\author{J. Neugebauer }
\affiliation{
Max-Planck-Institut f\"{u}r Eisenforschung GmbH, Max-Planck-Stra{\ss}e 1, 40237 D\"{u}sseldorf, Germany}%

\author{M. Himmerlich}
\email{marcel.himmerlich@tu-ilmenau.de}
\affiliation{Institut f{\"u}r Physik and Institut f{\"u}r Mikro- und Nanotechnologien MacroNano, Technische Universit{\"a}t Ilmenau, PF 100565, 98684 Ilmenau, Germany}%

\author{S. Krischok}
\affiliation{Institut f{\"u}r Physik and Institut f{\"u}r Mikro- und Nanotechnologien MacroNano, Technische Universit{\"a}t Ilmenau, PF 100565, 98684 Ilmenau, Germany}%

\author{M. Rink}
\affiliation{Institut f{\"u}r Physik and Institut f{\"u}r Mikro- und Nanotechnologien MacroNano, Technische Universit{\"a}t Ilmenau, PF 100565, 98684 Ilmenau, Germany}%

\author{J. Kr{\"o}ger}
\affiliation{Institut f{\"u}r Physik and Institut f{\"u}r Mikro- und Nanotechnologien MacroNano, Technische Universit{\"a}t Ilmenau, PF 100565, 98684 Ilmenau, Germany}%

\author{V. M. Polyakov}
\affiliation{Fraunhofer-Institut f{\"u}r Angewandte Festk{\"o}rperphysik, Tullastra{\ss}e 72, 79108 Freiburg, Germany}%

\date{\today}

\begin{abstract}
The adsorption of hydrogen at nonpolar GaN$(1\overline{1}00)$ surfaces and its impact on the electronic and vibrational properties is investigated using surface electron spectroscopy in combination with density functional theory (DFT) calculations. For the surface mediated dissociation of H$_2$ and the subsequent adsorption of H, an energy barrier of 0.55\,eV has to be overcome. The calculated kinetic surface phase diagram indicates that the reaction is kinetically hindered at low pressures and low temperatures. At higher temperatures ab-initio thermodynamics show, that the H--free surface is energetically favored. To validate these theoretical predictions experiments at room temperature and under ultrahigh vacuum conditions were performed. They reveal that molecular hydrogen does not dissociatively adsorb at the GaN$(1\overline{1}00)$ surface. Only activated atomic hydrogen atoms attach to the surface. At temperatures above 820\,K, the attached hydrogen gets desorbed. The adsorbed hydrogen atoms saturate the dangling bonds of the gallium and nitrogen surface atoms and result in an inversion of the Ga--N surface dimer buckling. The signatures of the Ga--H and N--H vibrational modes on the H-covered surface have experimentally been identified  and are in good agreement with the DFT calculations of the surface phonon modes. Both theory and experiment show that H adsorption results in a removal of occupied and unoccupied intragap electron states of the clean GaN$(1\overline{1}00)$ surface and a reduction of the surface upward band bending by 0.4\,eV. The latter mechanism largely reduces surface electron depletion.
\end{abstract}

\maketitle

\section{INTRODUCTION}\label{Int-GaN}

Gallium nitride (GaN) based optoelectronic devices are well established in solid state lighting \cite{Pimputkar2009,Crawford2009,Chang2012} and power electronics.\cite{Mishra2008,Ikeda2010,Chowdhury2015} High electron mobility transistor (HEMT) structures based on AlGaN/GaN have for example been modified in order to use these polar thin film devices as sensors with open or functionalized gate \cite{steinhoff03,pear04}, which essentially consist of the bare surface in polar orientation interacting with the surrounding species. In recent years, three-dimensional GaN nanowire structures have also attracted much attention for sensing applications \cite{chen2009,Ganguly2009,pearton2010,Teubert2011}, light emission and detection \cite{Guo2010,li2012,gonzales2012} as well as solar water splitting \cite{Shen10JPC,Wang2011,Wang12JPC} and photovoltaics.\cite{Tang2008,Dong2009} They exhibit a high surface to volume ratio and are mainly composed of side facets consisting of the nonpolar $m$-plane $(1\overline{1}00)$ surface. Such nanowire structures are typically grown by catalyst-free molecular beam epitaxy \cite{Wang2006Nanotech,Songmuang2007} and exhibit superior structural quality being almost free of defects and strain \cite{Schlager2008,cheze2010} with the capability to integrate vertical core-shell \cite{Qian2008,Koester2011,Yeh2012} and embedded lateral heterostructures \cite{Nguyen2011,Kehagias2013} and to intentionally dope the material.\cite{Guo2010,Schoermann2013}

For sensor applications as well as to identify optimum growth conditions a detailed understanding of the interaction and adsorption of molecules and atoms in the gas phase with the surface is crucial. Of special importance is also the existence of free or saturated bonds and related surface, adsorbate or trap states with well defined electronic structure.\cite{Eller2013} They can induce charge transfer processes, band bending including accumulation or depletion of electrons and/or form a dipole at the surface or interface in focus.\cite{Zhang2012,Robertson2013}  For example, the characteristics of GaN single-nanowire transistors have been found to be dependent on the valence band bending at the $m$-plane side facets which is directly influenced by surface adsorbates.\cite{san2013}

Hydrogen is a simple model adsorbate system and is known to affect doping in GaN.\cite{Neugebauer1996,Pearton1999,Neugebauer1999} Understanding its influence is of great technological relevance, since GaN bulk and thin film growth techniques involve hydrogen directly or indirectly as possible dissociation product from ammonia or metalorganic precursors with impact on the growth and properties of the resulting material.\cite{Ambacher1997,Okamoto1999,Aujol2001} For polar GaN surfaces, the interaction with hydrogen has been studied both by experimental methods \cite{Chiang1995,Shekhar1997,bell99,bell99_b,slo99,gra2000,star00} and theory.\cite{Pignedoli2001,Walle2002,Northrup2004,Bermudez2004,Chen2010,Kempisty2011,Kempisty2012,Ptasinska2015}

The nonpolar $m$-plane GaN$(1\overline{1}00)$ orientation is a low energy surface of wurtzite GaN \cite{Dreyer2014} and consists of buckled Ga--N dimers in the outermost layer.\cite{Nor96-PhysRevB.53.R10477,Segev2007L15} The dangling bonds at the nitrogen and gallium surface atoms give rise to occupied and unoccupied surface states that have been predicted by density functional theory (DFT) calculations \cite{seg06,ber09,Lym_APL2013} and experimentally verified.\cite{Lym_APL2013,him_APL2014} The chemically clean surface has been experimentally observed to exhibit a distinct electron depletion layer with a surface upward band bending of $\sim$\,0.6\,eV. The position changes in the presence of gas molecules that adsorb on the surface.\cite{him_APL2014} The kinetics and thermodynamics of adsorption and desorption, the resulting surface/adatom geometric structures and their influence on the electronic properties of the surface and subsurface region are crucial to understand charge transfer processes across the semiconductor/adlayer interface as well as ionized adsorbate induced formation of surface dipoles.

DFT studies have focused on the stability of hydrogen species at nonpolar $m$-plane \cite{nor97} and $a$-plane \cite{Akiyama2011} GaN surfaces and found that under hydrogen-rich conditions  hydrogen adsorbates attach at the free surface dangling bonds of both the Ga and N dimer atoms. 
It was further predicted that water molecules interacting with this surface spontaneously dissociate and form H and OH that bond with the surface.\cite{shen2009} In this study we combine first-principles calculations with surface adsorption/desorption experiments to clarify the mechanisms by which H and H$_2$ adsorb onto this surface. Based on this insight we study how the adsorbed H atoms modify the vibrational and electronic properties of this surface. We show that these aspects have consequences on growing GaN in hydrogen-rich environments and derive consequences when using nonpolar surfaces in electronic and chemical sensing devices.

\section{EXPERIMENT AND THEORY}\label{Det-GaN}
Clean GaN$(1\overline{1}00)$ surfaces were prepared by homoepitaxial overgrowth on bulk GaN substrates from Kyma Technologies produced by hydride vapor phase epitaxy using molecular beam epitaxy (MBE). The used substrates were unintentionally $n$-doped crystals with a carrier concentration of $\sim$\,5$\times$10\textsuperscript{16}\,cm\textsuperscript{-3} and had an epi-ready surface finish achieved by a final chemo-mechanical polishing step.\cite{pas09} Atomic force microscopy measurements identify atomically flat surfaces exhibiting a terrace width in correspondence with the sample miscut and a root-mean-square roughness below 0.5\,nm. Growth of an a few hundred nanometer thick GaN epilayer was performed using a Knudsen cell for Ga evaporation and a SVTA RF 4.5 plasma source (13.56 MHz) for the generation of reactive nitrogen species. The growth parameters Ga flux and substrate temperature were optimized at a constant nitrogen flux ($p_N$\,$\sim$\,5$\times$10\textsuperscript{-8}\,bar, plasma power 450\,W) using reflection high energy electron diffraction (RHEED) during, and photoelectron spectroscopy (PES) after growth, in order to obtain stoichiometric surfaces which are free of excess Ga or surface defects. The properties of the GaN$(1\overline{1}00)$ samples after growth have been reported and discussed earlier in Ref.\,\onlinecite{him_APL2014}. After epitaxy and cooling down, the samples were directly transferred under ultrahigh vacuum (UHV) conditions (base pressure $<$\,2$\times$10\textsuperscript{-13}\,bar) to the respective position for \textit{in-situ} surface analysis by ultraviolet and X-ray photoelectron spectroscopy (UPS, XPS). These measurements were performed in normal emission using a hemispherical electron analyzer and monochromated AlK$\alpha$ (1486.7\,eV) or He\,I (21.2\,eV) radiation for electron excitation. The description of the employed experimental conditions and parameters for PES can be found in Ref.\,\onlinecite{him07}.

To investigate their interaction with hydrogen, the \textit{as-grown} samples were exposed at room temperature to hydrogen (purity 99.999\,\%) by backfilling the analysis chamber (p$ _{H_{2}} $\,=\,2.0\,$ \times $\,10$ ^{-11}$\,bar) for up to 55\,min. Prior to each adsorption experiment, the gas supply lines were thoroughly evacuated to a pressure below 1\,$\times$\,10$^{-10}$\,bar and subsequently filled with 1.5\,bar H$_2$. The molecular hydrogen was optionally activated by a hot filament placed close to the sample front side to obtain atomic hydrogen by partial dissociation of the H$_2$ molecules in front of the GaN surface. Due to a limited cracking efficiency of the hot filament, the actual amount of produced atomic H species is below the calculated total H$_2$ exposure. During exposure, the residual gas was monitored by quadrupole mass spectrometry to control gas purity and absence of impurities. The pressure was measured with a Bayard Alpert ionization gauge and used without any further correction of specific gas sensitivity to calculate the exposure in Langmuir (1L\,=\,1.33$ \times $\,10$ ^{-9}$\,bar$\cdot$s). The changes of the surface properties were examined in-situ by UPS and XPS.

In an additional experiment, the hydrogen-covered $m$-plane GaN sample was transferred to a second UHV recipient using a vacuum transfer chamber with a base pressure $<$\,1$\times$10\textsuperscript{-11}\,bar. While clean surfaces were found to be extremely reactive to molecules from the residual gas, hydrogen adsorption at the GaN$(1\overline{1}00)$ surface resulted in a passivation of the surface and relatively stable conditions for vacuum transfer. The second UHV recipient is optimized for electron energy loss spectroscopy (EELS) using an Ibach spectrometer.\cite{IBACH1993819} Such experiments were performed on the H-covered GaN$(1\overline{1}00)$ surface in specular scattering geometry with energies of monochromatic electron beams varying between 5 and 80\,eV and were repeated after desorption of the hydrogen adsorbates by annealing the sample in UHV at 820\,$\pm$\,50\,K.

The H adsorption energies have been calculated employing DFT, the generalized gradient approximation (GGA) and the projector augmented-wave (PAW) method.\cite{Kresse93,Kresse96} The Ga 3$d$ electrons are included as valence states. The surfaces are modeled using a slab geometry consisting of 12 Ga-N monolayers (MLs) and a vacuum region of 20~\AA. A plane-wave energy cutoff of 450~eV was used and the Brillouin zone (BZ) was sampled using an equivalent 4\,$\times$\,4\,$\times$\,1 Monkhorst-Pack $k$-point mesh for the 1\,$\times$\,1 surface unit cell. The lowermost Ga and N atoms were passivated with pseudohydrogen having a fractional charge of 0.75 and 1.25, respectively. Convergence with respect to $k$-point sampling, energy cutoff, vacuum, and slab thickness were explicitly tested and found to provide surface energies with an accuracy better than 3\,meV/$1\times1$.

In order to investigate the vibrational properties and the vibrational entropic contributions we have calculated the dynamical matrix of the free and adsorbate covered surfaces. The force constant matrix and in turn the dynamical matrix have been calculated for the top four surface atomic layers and the H atoms at the surface using slabs of 8 layer thickness, 4\,$\times$\,4 surface supercells and a displacement of 0.01~\AA\, in both directions. To evaluate the H$_2$ chemical potential we have included the translational, rotational, and vibrational contributions. More details regarding the approach and the convergence criteria can be found in Ref.\,\onlinecite{Duff2014}.

The kinetics of dissociative adsorption of molecular H$_2$ are addressed with the harmonic transition state theory.\cite{Vineyard1957} The transition states have been identified by climbing image nudged elastic band (NEB) calculations.\cite{NEB} For the NEB calculations a 2\,$\times$\,2 surface slab with a thickness of 8 MLs has been implemented and in total 6 images including the two stable/metastable end states, i.e. H$_2$ in the vacuum and H$_2$ bound to a surface dimer, have been used to identify the transition points. The electronic structure of clean and H-covered $m$-plane surfaces have been computed with the Heyd, Scuseria, and Ernzerhof hybrid functional.\cite{HSE} This functional gives a bulk bandgap of $E_g$\,=\,3.116\,eV in agreement with previous calculations.\cite{RN2097}

\section{RESULTS AND DISCUSSION}\label{Res-GaN}

\subsection{Monitoring of hydrogen adsorption/desorption}\label{Res-Elec clean}
After growth the clean GaN$(1\overline{1}00)$ surface exhibits electron depletion with strong upward band bending. A distinct occupied surface state is found close to the valence band edge at 3.1\,eV below the Fermi energy. This feature was identified as emission from an occupied surface state related to the filled dangling bond states located at the N atoms of the GaN$(1\overline{1}00)$ surface dimer structure. For a detailed discussion on the electronic properties after growth we refer to our earlier study.\cite{him_APL2014} Since the focus of the present study is on hydrogen adsorption the variation of the valence band (VB) spectra has been monitored using UPS in continuing hydrogen interaction experiments increasing the exposure up to 50 Langmuir\,(L). In a first experiment molecular H$_2$ was offered to the clean GaN$(1\overline{1}00)$ surface at room temperature (RT) resulting in negligible changes of the valence band features and surface band bending (not shown). Consequently, at RT no significant H$_2$ dissociation and H adsorption is observed for the used H$_2$ partial pressure of 2.0\,$ \times $\,10$ ^{-11} $\,bar and the chosen reaction time. 

\begin{figure}[t]
\includegraphics[width=1\columnwidth]{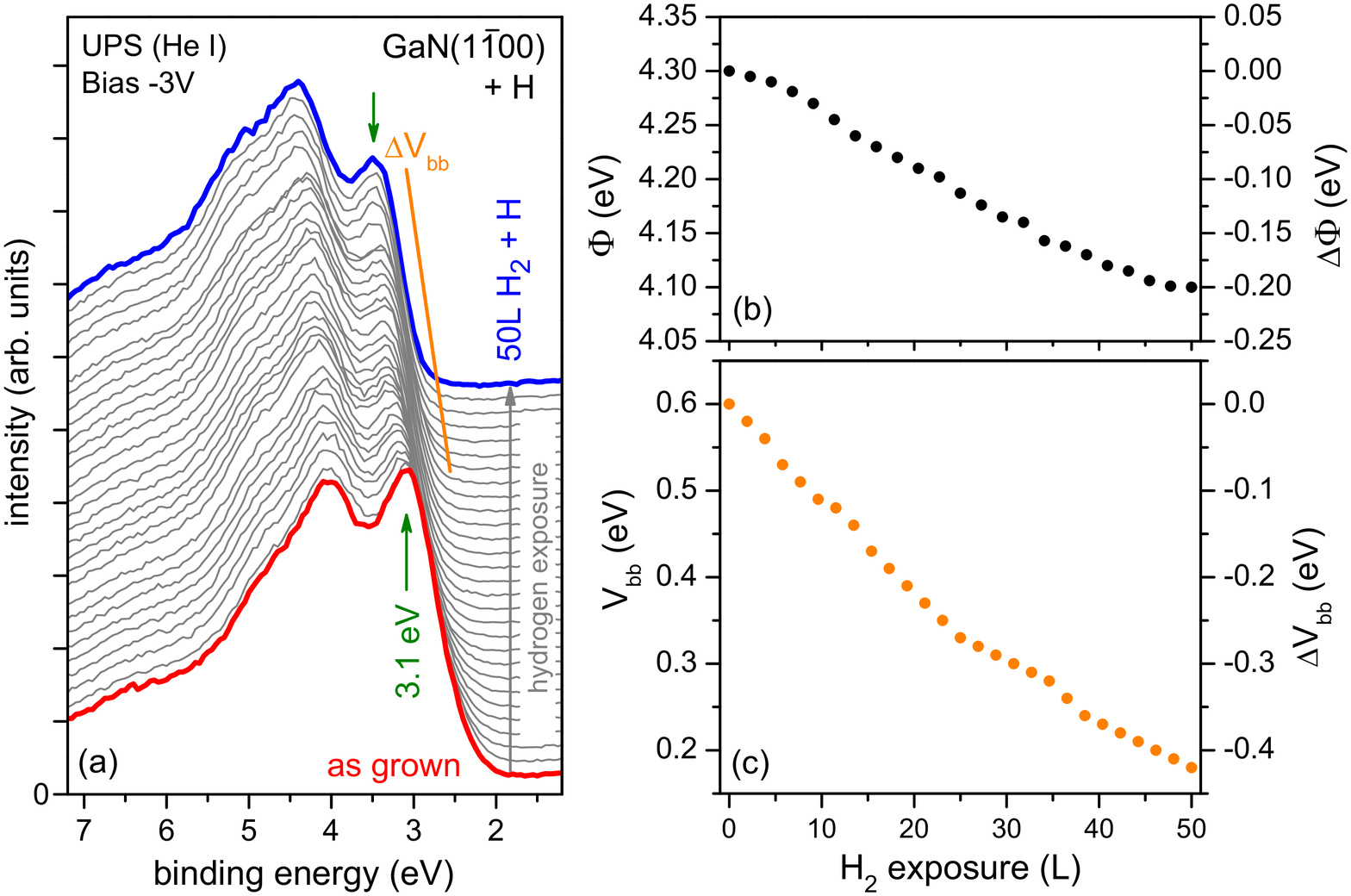}
\caption{\label{fig:fig_exposure} Changes in the GaN$(1\overline{1}00)$ surface electronic properties during continuous adsorption of atomic hydrogen (H) produced by a hot filament in the presence of H$_2$ -- (a) UPS (He\,I) valence band spectra revealing a shift of the occupied states and a reduction of electron emission from the surface state at 3.1\,eV. (b) Change of work function and (c) reduction in surface band bending in dependence upon hydrogen exposure.}
\end{figure}

\begin{figure}[t]
\includegraphics[width=0.8\columnwidth]{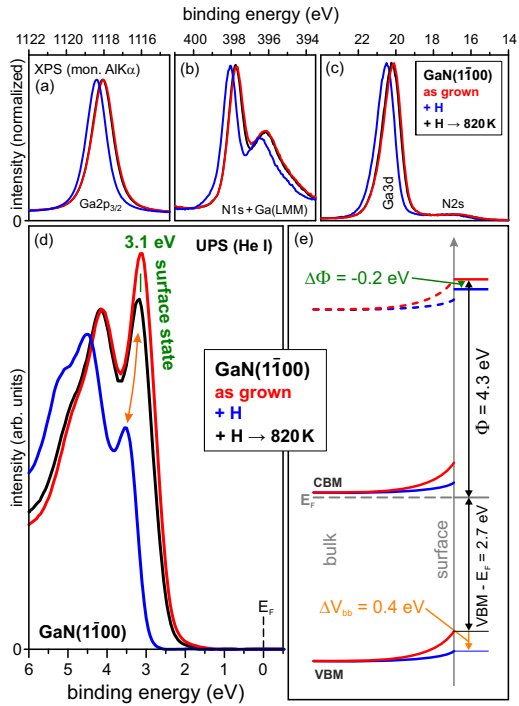}
\caption{\label{fig:fig_PES} Comparison of the GaN$(1\overline{1}00)$ surface properties after growth, atomic hydrogen adsorption and subsequent annealing at 820\,$\pm$\,50\,K. (a)\,--\,(c) X-ray photoelectron spectra of the Ga2p$_{3/2}$ and Ga3d states as well as the N1s core level including contributions from the Ga(LMM) Auger transition. The individual core level spectra were normalized with respect to their maximum peak height. (d) Valence band structure measured by UPS using He\,I radiation. A subtraction of contributions from He\,I satellite lines was applied. (e) Electronic properties including work function and surface band alignment of the m-plane GaN surface with and without adsorbed hydrogen determined based on the results of the photoelectron spectroscopy measurements.}
\end{figure}

As a consequence, in a following experiment we have initiated the H adsorption process by implementing a hot filament for partial generation of thermally activated  atomic hydrogen close to the sample surface. Fig.\,\ref{fig:fig_exposure}\,(a) shows a series of UPS (He\,I) valence band spectra during ongoing H and H$_2$ exposure up to 50\,L in total. A gradual shift of the occupied states away from the Fermi level $E_F$ at 0\,eV is observed. This effect is directly linked to a reduction of the surface band bending, which initially amounts to 0.6\,eV for the as grown $m$-plane GaN surface.\cite{him_APL2014} In addition, the onset of low-energy secondary electron emission (not shown) was shifted, pointing to a reduction of the work function $\Phi$. The temporal variation of $\Phi$ as well as the determined change in surface band bending $\Delta V_{bb}$ are plotted in Fig.\,\ref{fig:fig_exposure}\,(b) and (c), respectively. Both values decrease monotonically with H exposure, with the tendency of saturation at the end of the experiment. In parallel, the signal intensity of the filled N dangling bond state, initially found at 3.1\,eV binding energy, is significantly reduced upon the interaction process [Figs.\,\ref{fig:fig_exposure}\,(a) and \ref{fig:fig_PES}\,(d)]. These aspects provide indirect evidence for H adsorption at the GaN$(1\overline{1}00)$ surface for this second experiment in which the sample was exposed to activated H species.

To prove the observed shift in surface band bending, we have also characterized the core level binding energies using AlK$\alpha$ X-ray excitation. The corresponding spectra of the Ga2p$_{3/2}$, N1s and Ga3d states are shown in Fig.\,\ref{fig:fig_PES}\,(a)\,--\,(c). Their respective binding energies for the as-grown surface are 1118.0, 397.7 and 20.1\,eV. We emphasize that no surface impurities were detected by XPS after the MBE growth or the subsequent H$_2$ exposure, and therefore the observed changes are not induced by surface impurities, while unfortunately a direct detection and analysis of hydrogen surface species is not possible by XPS. After the performed H exposure, in all cases a shift of the core level binding energy by 0.4\,eV towards higher values is observed, consistent with the changes observed in the He\,I spectra during the adsorption process. These observations indicate that the reaction with the activated hydrogen species results in a saturation of the free dangling bonds at the Ga--N surface. The impact of the saturation will be discussed in detail in comparison with theoretical predictions below.

In order to analyze the stability or reversibility of the H adsorption process, we have afterwards annealed the sample in ultrahigh vacuum. While an annealing temperature of 520\,$\pm$\,50\,K did not substantially change the surface electronic properties in terms of band bending and work function indicating a certain stability of the adsorbate structure, heating the substrate up to 820\,$\pm$\,50\,K almost recovered the characteristics after MBE growth. Figs.\,\ref{fig:fig_PES}\,(a)\,--\,(d) include the photoelectron spectra obtained after annealing the hydrogenated surface at 820\,$\pm$\,50\,K for 10\,min (black). Obviously, the occupied states shift back towards the Fermi level and most noticeably, the near VB edge emission from the occupied surface state also recovers in intensity to almost the signal strength after growth [Fig.\,\ref{fig:fig_PES}\,(d)]. Consequently, the hydrogen atoms adsorbed with an effective coverage in the submonolayer regime during the reaction of activated H species and the clean GaN$(1\overline{1}00)$ surface, can be reversed, i.e. desorption initiated, if sufficient energy is introduced into the system, e.g. thermally induced as examined in this experiment.

Fig.\,\ref{fig:fig_PES}\,(e) schematically summarizes the experimentally determined differences in electronic properties for the clean and H-covered GaN$(1\overline{1}00)$ surface including variation in surface band bending and work function.

The aforementioned results indicate a complex interplay between thermodynamics and kinetics in hydrogen adsorption/desorption processes: they imply that it is energetically favorable for atomic hydrogen to adsorb at the surface, passivate the Ga-- and N-- dangling bonds of the clean GaN$(1\overline{1}00)$ surface and saturate the surface states. This results in a reduction of the upward band bending by 0.4\,eV and reduction of the work function by 0.2\,eV as illustrated in the surface band diagram in Fig.\,\ref{fig:fig_PES}\,(e). However, hydrogen desorption can be mediated at elevated temperatures which can be attributed to the existence of a kinetic barrier. These aspects will be further addressed below.

\subsection{Electron density profile and surface vibrations}

\begin{figure}[t]
\includegraphics[width=1\columnwidth]{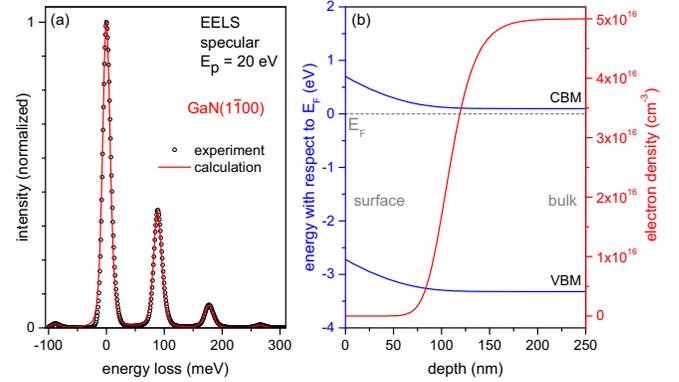}
\caption{\label{fig:fig_EELS_Sim} (a) Specular EEL spectra (circles) of clean GaN$(1\overline{1}00)$.  The energy of incident electrons was set to 20\,eV.  The full line depicts calculated results.  The prominent loss features are due to single and multiple electron scattering from the Fuchs-Kliewer phonon at 88\,meV. (b) Surface band alignment and electron density profile resulting from modeling the experimental EEL spectra measured using different primary electron energies. }
\end{figure}

While PES unraveled changes in the electronic structure upon H adsorption, vibrational spectroscopy was performed to confirm adsorption of atomic H and to identify adsorption sites. The discovered possibility to saturate the surface by atomic hydrogen, forming a stable adsorbate (passivation) layer was used to transfer the H-covered GaN samples to a separate UHV chamber. The same heating procedure as used for the PES analysis was then performed to remove the hydrogen adatoms and to analyze the vibrational characteristics of the GaN$(1\overline{1}00)$ surface.

Figure\,\ref{fig:fig_EELS_Sim}\,(a) shows a representative spectrum measured after transfer of the H-covered GaN sample and annealing at 820\,$\pm$\,50\,K in specular scattering geometry and a primary electron energy ($E_p$) of 20\,eV. Apart from the signature of elastically scattered electrons at 0\,meV the spectra of the clean surface exhibit almost equidistantly separated loss features.  These peaks are assigned to the spectroscopic signatures of the Fuchs-Kliewer (FK) phonon at (88.0\,$\pm$\,0.4)\,meV (average value of EEL spectra measured at different $E_p$) and its multiple excitations.  This interpretation is corroborated by previous findings for GaN(0001) and GaN$(000\overline{1})$ surfaces \cite{slo99, bell99, gra2000} and the simulations to be discussed next.

EELS data were simulated using a model which is similar to the theoretical approach developed in Refs.\,\onlinecite{lam85,lam90}. In these calculations, the surface energy-loss function is derived using the continued-fraction expansion method \cite{lam90}, when the subsurface region is represented by a finite number of sublayers of certain thickness to reproduce a smooth variation of the depth dependent electron density below the surface. For this purpose, electron density depth profiles are computed by solving the Schr\"odinger and Poisson equations self-consistently. It should be noted that the model used for fitting the measured spectra includes only two contributions, originating in collective lattice vibrations (phonons) and oscillations of the free-electron gas in the conduction band (plasmons). Also, due to sufficiently low bulk electron density [see Fig.\,\ref{fig:fig_EELS_Sim}\,(b)] in these samples, the plasmons can manifest themselves only as a small broadening of the elastic peak, in addition to the instrumental resolution ($\sim$\,3\,meV for the implemented experimental conditions).

The resulting electron density profile from modeling the combined EEL spectral data with varying $E_p$ and the associated band edge alignment in the near surface region are depicted in Fig.\,\ref{fig:fig_EELS_Sim}\,(b). The calculations are in very good agreement with the qualitative model of upward band bending [Fig.\,\ref{fig:fig_PES}\,(e)] with an $V_{bb}$ value of 0.7\,eV compared to 0.6\,eV extracted from the PES measurements. Furthermore some important quantitative information can be extracted: the calculations indicate a bulk electron concentration of 5\,$\times$\,10$^{16} $\,cm$^{-3}$ combined with a depletion layer thickness of $\sim$\,100\,nm. Consequently, these results confirm the low electron concentration in these samples and that PES can be used for the determination of the band bending values since the width of the depletion layer is much larger than the information depth of a few nanometers from the surface obtained in PES. In this region the slope of the bands is found to be negligible within the uncertainty of 0.1\,eV.

\begin{figure}[t]
\includegraphics[width=0.9\columnwidth]{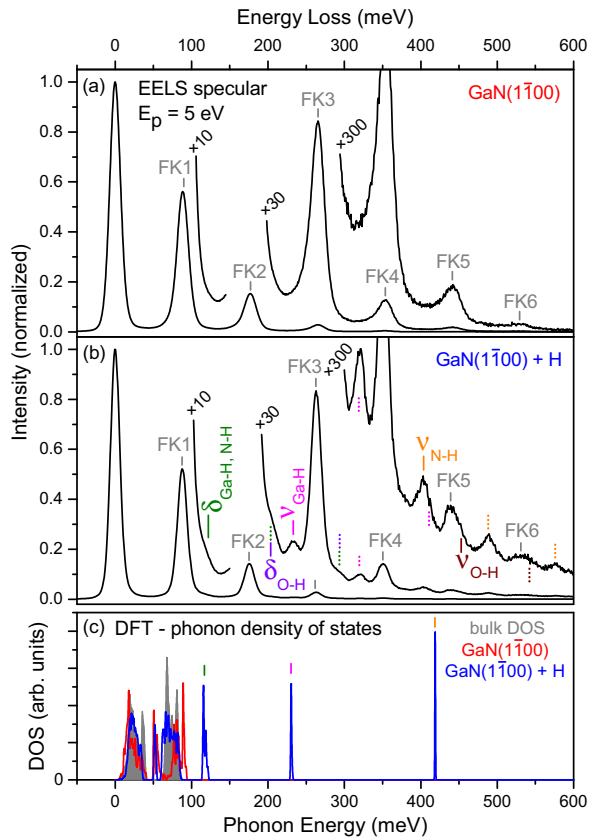}
\caption{\label{fig:fig_EELS} (a) Specular EEL spectrum of clean GaN$(1\overline{1}00)$ with FK$n$ indicating loss features that result from single ($n$\,=\,1) and multiple ($n$\,$\geq$\,2) electron scattering from the Fuchs-Kliewer (FK) phonon at 88\,meV.  The incident electron energy was set to 5\,eV.  (b) As (a) for H-covered GaN$(1\overline{1}00)$. (c) Calculated phonon density of states (DOS) of clean (red) and H-covered (blue) GaN$(1\overline{1}00)$ surfaces.  The gray shaded area depicts the projected bulk phonon DOS. Additional loss features in (b) and vibrational modes in (c) are due to Ga--H and N--H bending modes ($\delta_{\text{Ga--H}}$ and $\delta_{\text{N--H}}$) with calculated energies around 116\,meV. Ga--H and N--H stretching vibrations exhibit experimental\,(calculated) vibrational energies of $\nu_{\text{Ga--H}}$\,=\,233\,($\sim$\,231)\,meV and $\nu_{\text{N--H}}$\,=\,403\,(419)\,meV, respectively. The O--H bending mode appears at $\delta_{\text{O--H}}$\,=\,206\,meV and the detected O--H stretch mode has an energy of $\nu_{\text{O--H}}$\,=\,453\,meV. The surface excitations are indicated and related FK combination losses are marked using dotted lines of the same color.}
\end{figure}

Figure \ref{fig:fig_EELS} shows specular EEL spectra of clean [Fig.\,\ref{fig:fig_EELS}\,(a)] and H-covered [Fig.\,\ref{fig:fig_EELS}\,(b)] GaN$(1\overline{1}00)$ surfaces that were acquired with an incident electron energy of 5\,eV.  The equidistant cascade of loss features results from the single (FK1) and the multiple (FK2\,--\,FK6) excitation of the FK phonon. The single surface phonon energy (FK1) amounts to 88\,meV, slightly higher than those reported for GaN(0001) and $(000\overline{1})$ surfaces.\cite{slo99, bell99, gra2000} For the H-covered surface, additional signatures are observed and marked by solid lines in Fig.\,\ref{fig:fig_EELS}. Corresponding FK phonon combination losses (e.g. $\nu_{\text{{Ga--H}}}$\,+\,FK1) are marked by dotted lines of the same color. The feature at 233\,meV is caused by Ga--H stretching vibrations \cite{gra2000} of H adsorbed at the Ga-- dangling bond of the Ga--N dimer. The corresponding N--H stretching vibration mode is also observed at 403\,meV.

In Fig.\,\ref{fig:fig_EELS}\,(c) the projected bulk phonon density of states (DOS) (gray shaded area) as well as the phonon DOS of the clean and H-covered GaN$(1\overline{1}00)$ surfaces (red and blue curves, respectively) are shown. Each DOS is the sum over all states within the respective bulk or surface Brillouin zone. Besides the acoustic (0\,--\,40\,meV) and optical (60\,--\,90\,meV) bulk phonon modes \cite{PhysRevB.58.12899}, which are not directly detected in the EELS experiment, three additional features are found for the m-plane surface with H atoms adsorbed at the Ga--N dimer dangling bonds.
The N--H stretching mode ($\nu_{\text{N--H}}$) is identified as nondispersing state at a calculated vibrational energy of 419\,meV and the corresponding Ga--H stretching mode ($\nu_{\text{Ga--H}}$) exhibits slight dispersion between 229 and 232\,meV within the BZ (231.4\,meV at the $\Gamma$-point). Both values are in fairly good agreement with the EELS experiment. In addition, the signature between 113\,meV and 123\,meV can be assigned to bending vibrations of H atoms adsorbed at the surface dimer structure ($\delta_{\text{Ga--H}}$ and $\delta_{\text{N--H}}$).\cite{nor97} The calculations reveal that this structure consists of two states that disperse in the BZ within the mentioned energy range with a $\Gamma$-point energy of 115.6 and 117.2\,meV, respectively. These vibrational energies are nearly two times larger than the energy reported for the Ga--H bending mode at the GaAs(110) surface.\cite{gra96} The deviation may be ascribed to the difference of the microscopic adsorption geometry.

A faint shoulder on the high-energy side of the first FK phonon loss feature is observed around 118\,meV and is attributed to these calculated Ga--H and N--H bending modes, rather than to N--OH vibrations that were previously reported to exist at a slightly lower vibrational energy of 106\,meV.\cite{gra2000} The additional shoulder at 206\,meV is attributed to a superposition of contributions from a FK phonon combination loss ($\delta_{\text{{Ga--H,N--H}}}$\,+\,FK1) and an O--H bending vibrational mode ($\delta_{\text{{O--H}}}$).\cite{gra2000} The latter aspect is corroborated by the presence of a weak loss structure at 453\,meV as side feature of the FK5 multiple, which is caused by O--H stretching vibrations ($\nu_{\text{{O--H}}}$).\cite{slo99,gra2000} Consequently, the main spectroscopic features are assigned to vibrations of atomic H adsorbed to Ga and N surface atoms.  A slight uptake of hydroxides is indicated by the EEL spectra and is due to the high reactivity of unsaturated GaN surfaces \cite{star00} combined with the two orders of magnitude higher base pressure in the used vacuum transfer system compared to the recipient for in-situ PES analyses.  After annealing at 820\,$\pm$\,50\,K signatures of adsorbate vibrational modes fall below the detection limit of the spectrometer [Fig.\,\ref{fig:fig_EELS}\,(a)]. As a result, the EELS experiment provided important information that atomic H saturates Ga-- and N-- dangling bonds of the surface dimer structure.

\subsection{Influence of hydrogen on the structural and electronic properties of GaN$(1\overline{1}00)$ surfaces}\label{Res-Elec hydrogen}

In order to investigate the effect of H adsorption on the electronic properties of GaN surfaces and to develop a microscopic model of the differences in surface geometry, we have performed DFT calculations to compute the surface crystal structures and the band structures of the clean and hydrogen-covered $m$-plane GaN surfaces. In Table~\ref{tbl:tbl} the PBE-GGA and HSE calculated displacements from the bulk like positions of the Ga and N surface atoms as well as the buckling angles of the surface cation--anion dimers at clean and hydrogen-covered GaN$(1\overline{1}00)$ surfaces are shown. After structure relaxation, the cations (Ga atoms) at the clean surface move inwards adopting an $sp^2$-like configuration and the anions (N atoms) move outwards in an $sp^3$-like configuration. Relaxation results in  $\approx$7.5\% contraction and $\approx8.1^\circ$ buckling angle of the Ga--N bond. Furthermore, the back bond length between the surface Ga\,(N) atoms and the N\,(Ga) atoms in the first subsurface layer is contracted by 2.79\% (3.55\%).

\begin{table}[t]
  \caption{Atomic displacements from their bulk-like positions in \AA\, and buckling angles $\omega$ (with respect to a virtual flat and symmetric surface dimer) of the top layer Ga and N atoms at the GaN$(1\overline{1}00)$ clean and H-covered surfaces. $\Delta x$, $\Delta y$, $\Delta z$ indicate displacements along [0001], $[11\overline{2}0]$, and $[1\overline{1}00]$ directions, respectively (see insets in Fig.\,\ref{fig:fig_band}). $\Delta r$ is the length of the displacement vector.}\label{tbl:tbl}

  \begin{ruledtabular}
\begin{tabular}{lccccc}

 atom & $\Delta x$ & $\Delta y$ & $\Delta z$ & $\Delta r$ & $\omega$\\
\hline
\multicolumn{6}{l}{Clean surface} \\
\hline
\multicolumn{6}{l}{PBE-GGA} \\
  \hline
  Ga & 0.16 & 0.00 & -0.28 & 0.32 & \multirow{2}{*}{$8.08^\circ$}\\
  N & -0.01 & 0.00 & -0.02 & 0.02  \\
  \hline
  \multicolumn{6}{l}{HSE} \\
  \hline
  Ga & 0.15 & 0.00 & -0.28 & 0.32 & \multirow{2}{*}{$8.14^\circ$}\\
  N & -0.01 & 0.00 & -0.02 & 0.02  & \\
  \hline
  \hline
  \multicolumn{6}{l}{Hydrogen-covered surface} \\
  \hline
   \multicolumn{6}{l}{PBE-GGA} \\
    \hline
  Ga & -0.05 & 0.00 & 0.07 & 0.09 & \multirow{2}{*}{-$3.90^\circ$}\\
  N & -0.03 & 0.00 &  -0.07 & 0.08 & \\
    \hline
    \multicolumn{6}{l}{HSE} \\
    \hline
    Ga & -0.06 & 0.00 & 0.06 & 0.09 & \multirow{2}{*}{-$3.76^\circ$}\\
    N & -0.05 & 0.00 &  -0.07 & 0.08 & \\
\end{tabular}
\end{ruledtabular}
\end{table}

At the hydrogen-covered surface both Ga and N surface atoms move outwards and the Ga--N bond length is expanded by $\approx$1.2\% with a buckling angle of $\approx-3.90^\circ$, i.e. in the opposite direction with respect to the clean surface. The Ga--H and N--H bond lengths are 1.57~\AA\, and 1.03~\AA, respectively. The bonds of the Ga (N) atoms at the top most surface layer with the N (Ga) atoms at the second layer are expanded (contracted) by 0.42\% (0.40\%). Hence, after hydrogen adsorption the Ga and N surface atoms adopt more bulk-like positions.

\begin{figure}[t]
	\includegraphics[width=1.0\columnwidth]{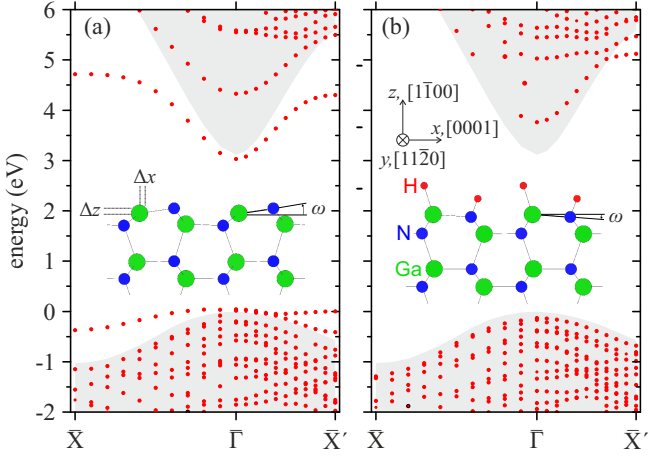}
	\caption{\label{fig:fig_band} Band structure of the (a) clean and (b) hydrogen-covered GaN$(1\overline{1}00)$ surfaces. The gray shaded areas indicate the projected bulk band structure. Insets: Ball and stick models of the corresponding surfaces in side view. The buckling angles of the Ga--N bonds $\omega$ are indicated. In (a) the displacements $\Delta x$ and $\Delta z$ of the Ga surface atoms from the bulk like positions are schematically shown.}
\end{figure}

The band structures of clean and hydrogen-covered GaN$(1\overline{1}00)$ surfaces are shown in Figs~\ref{fig:fig_band}(a) and (b), respectively. The clean $m$-plane GaN surface introduces a Ga-derived deep unoccupied $s$-type state at 2.98\,eV above the bulk valence band maximum (VBM) at the $\overline{\Gamma}$ point of the surface Brillouin zone. This value is larger than previous Hubbard-corrected LDA+U~\cite{Lym_APL2013} and PBE+U~\cite{PhysRevB.91.035302} calculations or specifically modified pseudopotential calculations~\cite{wal07} which yield a surface band gap of 2.4, 2.68 and 2.7\,eV, respectively. However, it is in good agreement with previous self-energy–corrected LDA-1/2 calculations which predict a surface band gap of 3.03\,eV but  smaller than the value of 3.31\,eV calculated by HSE with 32\% fraction of exact exchange and the Ga 3$d$ electrons treated as valence states.\cite{PhysRevB.91.035302} In all the aforementioned high level DFT calculations the unoccupied surface state is well below the bulk conduction band minimum (CBM). The differences in the calculated surface band gaps can be attributed to the different methods as well as to the different slab thickness employed in these calculations. However, the position of the unoccupied surface state, i.e. 2.98\,eV above the bulk VBM, is in good agreement with the measured band bending of $\approx0.6$\,eV as discussed in detail below.

The band structure of the hydrogen-passivated $m$-plane GaN surface is depicted in Fig.~\ref{fig:fig_band}(b): Passivation of the surface dangling bonds by hydrogen results in Ga--H and N--H occupied bonding and unoccupied antibonding states. The former shift from above the VB edge for the clean surface into the bulk VB region. The unoccupied states, initially also found as intragap states shift into the bulk conduction band (CB) region. Hence, hydrogen passivation shifts the surface states out of the fundamental band gap, providing a suitable explanation for the discovered changes in band bending as measured by PES. For the clean surface, the presence of deep unoccupied gap states induces a transfer of electrons from the CB into these energetically favorable localized states causing a depletion of the surface from electrons and strong upward band bending of 0.6\,--\,0.7\,eV as determined by PES and EELS simulations. The unoccupied states are pinning centers for the surface Fermi level as discussed in detail in Refs.\,\onlinecite{Lym_APL2013} and \onlinecite{him_APL2014}. If these states shift towards or even above the CB edge as calculated for the H-saturated surface, the surface Fermi level follows resulting in a reduced upward band bending $V_{bb}$ or even unpins the surface Fermi level resulting in flat band conditions. For the performed experiment, $\Delta V_{bb}$ is 0.4\,eV, indicating a remaining slight upward band bending/electron depletion. However, from the experimental data it is not possible to extract an exact number for the H-coverage for this experiment and one might expect an asymptotic convergence to the situation of a fully covered surface (considered in the calculations) for higher exposure.

\subsection{Thermodynamics and Kinetics of hydrogen adsorption}\label{Res-Ads}

\begin{figure}[t]
	\includegraphics[width=1.0\columnwidth]{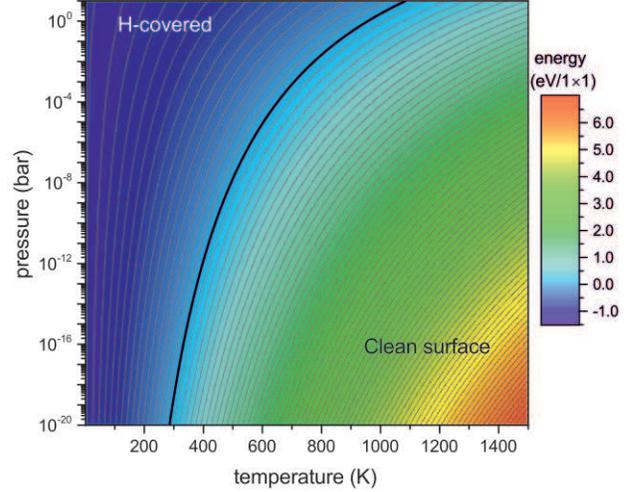}
	\caption{\label{fig:fig_thermod} Difference in the free energy, $\Delta F$ [Eq.\,(\ref{eq:thermodynamics1})], of the hydrogen-covered and clean GaN$(1\overline{1}00)$ surfaces as function of hydrogen pressure and temperature. The thick contour line indicates the range of pressures and temperatures where both systems are in equilibrium. Blue (red) colors indicate smaller (larger) values. Each contour line corresponds to an energy difference of 0.1\,eV per 1$\times$1 surface cell area. In the region to the left of the equilibrium line (thick black line) the hydrogen-covered surface is thermodynamically favored.}
\end{figure}

In order to address the thermodynamics of hydrogen adsorption, the free energy difference $\Delta F$ of the hydrogen-covered and clean GaN$(1\overline{1}00)$ surfaces was calculated as
\begin{equation}\label{eq:thermodynamics1}
  \Delta F = \Delta E_\mathrm{tot}+\Delta F_\mathrm{vib}-\mu_{H_2}
\end{equation}
where $\Delta E_\mathrm{tot}= E_\mathrm{tot}^{(1\overline{1}00):2H}-E_\mathrm{tot}^{(1\overline{1}00)}$ is the difference between the total energies of the hydrogen-covered and the clean surfaces, $\Delta F_\mathrm{vib}= F_\mathrm{vib}^{(1\overline{1}00):2H}-F_\mathrm{vib}^{(1\overline{1}00)}$ is the difference in
vibrational contributions to the free energy, and $\mu_{H_2}$ is the chemical potential of H$_2$.
In Fig.~\ref{fig:fig_thermod} the difference in the surface free energies is plotted as function of temperature and H$_2$ pressure. Higher temperatures and/or lower pressures favor the clean surface. This is attributed to the large translational entropic contributions H$_2$ molecules have in the gas phase at these conditions. On the other side, at low temperatures and/or high partial pressures the hydrogen-covered surface is thermodynamically favorable. More specifically, at 300\,K and for H$_2$ pressures larger than $5\times10^{-19}$~bar it is thermodynamically favorable to adsorb hydrogen at the $m$-plane GaN surface. However, this is in contrast to the experimental finding that at the same temperature and at 8 orders of magnitude higher pressure (i.e. $2.0\times10^{-11}$~bar) of molecular H$_2$, no significant hydrogen adsorption is observed. Furthermore, the annealing experiments indicate that dehydrogenation of the surface, within the time scale of the experiments, requires elevated temperatures as high as 820\,K. This further indicates that kinetic effects rather than the thermodynamic properties control the H coverage on the surfaces. In order to identify and investigate these mechanisms we next focus on the adsorption and desorption kinetics of both atomic and molecular hydrogen.

The adsorption of atomic hydrogen is barrierless. The desorption/binding energy $E_\mathrm{des}$ of atomic hydrogen is defined as:
\begin{equation}\label{eq:eq_adsdes}
  E_\mathrm{des}=E_\mathrm{surf:H}-E_\mathrm{surf}-E_\mathrm{H_{atom}},
\end{equation}
where $E_\mathrm{surf:H}$ and $E_\mathrm{surf}$ are the total energies of the surface with and without adsorbed hydrogen atom, respectively and $E_\mathrm{H_{atom}}$ is the total energy of a hydrogen atom.
In the calculation of the desorption energies different effects have to be considered: First the N--H bond is stronger than the Ga--H. Second, both unpassivated and doubly passivated surface dimers, i.e. both Ga and N atoms of the same dimer are passivated by hydrogen, obey the electron counting rule and do not introduce occupied states deep in the fundamental gap. On the contrary, passivation of only cation or anion dangling bonds of one dimer will result in the formation of fully or partially occupied states deep in the gap region. Furthermore, as has already been discussed, surface relaxation and re-hybridization effects result in different atomic geometries for the clean and doubly passivated surface dimers (see Table~\ref{tbl:tbl}).

As discussed above, bond enthalpies, electronic structure and surface strain are expected to strongly influence the desorption energies. Thus, they have to be explicitly considered by investigating different atomic hydrogen desorption scenarios. Hence, different desorption mechanisms have been calculated. These correspond to H desorption from cations and anions at doubly and singly passivated surface dimers in the limit of clean and fully covered surfaces. Our calculations reveal that H desorption from cations is energetically preferred to desorption from anions by at least 0.5\,eV. This value is considerably higher than $k_B T$ at 300\,K (0.026\,eV) or even at temperatures as high as 1000\,K (0.086\,eV). The corresponding desorption energies are $\approx$2.5\,eV and $\approx$4.8\,eV in the limit of a clean or fully covered surface, respectively.

\begin{table}
	\caption{Calculated adsorption (ads.) and desorption (des.) barriers of H$_2$ molecules at clean and hydrogen-covered m-plane GaN surface in eV.}\label{tbl:kinetics}
	
	\begin{ruledtabular}
		\begin{tabular}{l|cc|cc}
			&	\multicolumn{2}{c}{clean} &	\multicolumn{2}{c}{H-covered} \\
			\hline
			& ads. & des. & ads. & des. \\	
			\hline	
			H$_2$ at Ga--N dimer & 	0.55 & 2.30 & 0.62 & 2.30 \\		
			H$_2$ at 2 Ga atoms  & 	2.41 & 0.52 & 0.43 & 2.85 \\
			H$_2$ at 2 N atoms   & 	1.57 & 2.53 & 0.15 & 5.43 \\
		\end{tabular}
	\end{ruledtabular}
\end{table}

\begin{figure}[b]
\includegraphics[width=1.0\columnwidth]{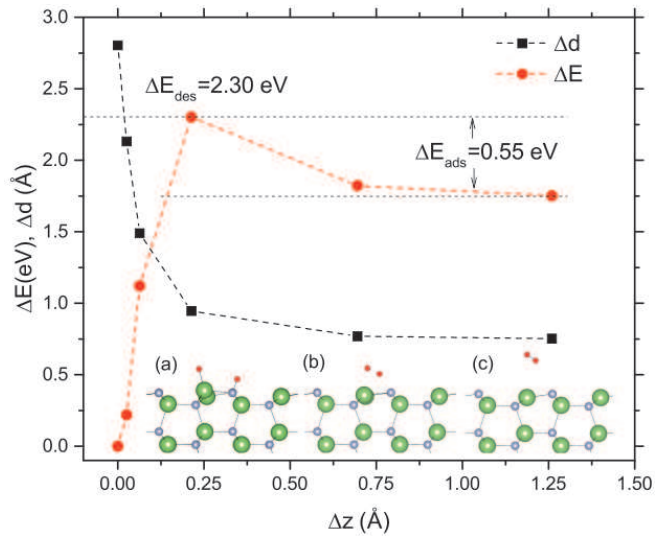}
\caption{\label{fig:fig_neb} Energy change $\Delta E$ and H--H interatomic distance $\Delta d$ along the minimum energy path for H$_2$ adsorption at the clean GaN$(1\overline{1}00)$ surface. The distance of the H$_2$ center of mass from the surface $\Delta z$ is used to represent the reaction coordinate. H$_2$ adsorbed at the surface is used as reference both for the energy change and the reaction coordinate. Insets: Schematic representation in side view along $[11\overline{2}0]$ of H$_2$ (a) bound to surface, (b) at the transition state, (c) and in the vacuum. Large green and smaller blue balls indicate Ga and N atoms, respectively. The H atoms are denoted by the smallest red spheres.}
\end{figure}

The aforementioned reaction mechanism might not be the most relevant since adsorption/desorption of hydrogen may be realized by the formation of H$_2$ molecules. As with atomic hydrogen desorption, different reaction mechanisms have been considered: Adsorption/desorption at (i) a surface dimer, (ii) two neighboring Ga surface atoms, and (iii) two neighboring N surface atoms considering the two limits, i.e. of a clean and fully covered surface. The corresponding energy barriers are listed in Table~\ref{tbl:kinetics}. The adsorption mechanism with the lowest adsorption barrier is H$_2$ dissociatively binding at a single surface dimer. In Fig.~\ref{fig:fig_neb} the energy change as well as the H--H interatomic distance along the minimum energy path for H$_2$ adsorption on a Ga--N dimer at a clean surface are plotted as function of the distance between the H$_2$ center of mass and the surface. The barrier for H$_2$ adsorption is $\approx0.55$~eV and corresponds to the energy required to dissociate the molecule. On the other hand the desorption energy is $\approx2.30$~eV. It should be noted that the aforementioned energy barriers depend weakly on the surface coverage. In the limit of a fully hydrogen-covered surface the adsorption and desorption energy barriers are $\approx0.62$~eV and $\approx2.30$~eV, respectively.

H$_2$ desorption from two neighboring N or Ga surface atoms in the limit of a fully covered surface has higher kinetic barriers, 5.43 and 2.85\,eV, respectively. These reaction mechanisms result in single passivated surface dimers and increase both the surface strain and the electronic contributions to the surface energy (see above). Interestingly, the desorption barrier of 0.52\,eV is remarkably rather small for H$_2$ binding at two neighboring Ga atoms in the limit of a clean surface. This is because it is highly unfavorable for H to passivate only surface cations even at extreme H-rich conditions. In order that this mechanism can actually take place, the H atoms would already have been desorbed from the N surface atoms. However, the latter has considerably higher kinetic barriers. Hence, desorption of molecular hydrogen from neighboring surface cations or anions can be neglected and molecular hydrogen adsorption and desorption is taking place by preferentially binding to and desorbing from Ga--N dimers.

The desorption barrier for the hydrogen molecule is considerably smaller than the desorption energy of atomic hydrogen. Hence, desorption is taking place as molecular H$_2$ rather than as atomic hydrogen. On the other hand for the adsorption of a H$_2$ molecule, an energy barrier has to be overcome, while adsorption of atomic hydrogen is barrierless. Thus, if both atomic and molecular hydrogen are present in the gas phase, then surface passivation by hydrogen will preferentially take place through atomic hydrogen adsorption and the rate limiting mechanism will be the flux of incident hydrogen atoms at the surface.

The flux of incident particles at a surface depends on the temperature and the corresponding partial pressure $p$ and is given by the Hertz-Knudsen equation:\cite{ANDP:ANDP18822531002,ANDP:ANDP19153521306}
\begin{equation}\label{eq:eqHK0}
  f\left(p,T\right)=\frac{p}{\sqrt{2\pi m k_B T}},
\end{equation}
where $m$ is the mass of the corresponding particles. For example, for an atomic hydrogen partial pressure of $10^{-11}$~bar at $T=300$~K the flux of incident atomic hydrogen is $\approx0.025$\,s$^{-1}$ per 1\,$\times$\,1 surface cell area. Under these conditions and assuming that hydrogen desorption is kinetically suppressed, 50\% or 100\% of a monolayer surface coverage of an initially clean surface will be achieved within $\approx 1$\,min and $\approx10$\,min, respectively. In contrast to atomic hydrogen the adsorption of H$_2$ molecules is not barrierless and the corresponding rate is given by the following equation:
\begin{equation}\label{eq:eqHK}
  \nu_\mathrm{ads}\left(p, T \right) = f\left(p,T\right) \cdot A \cdot \exp{\left(-\frac{E_\mathrm{ads}}{k_B T}\right)},
\end{equation}
where $A$ is the area of the 1\,$\times$\,1 surface unit cell and $E_\mathrm{ads}$ is the kinetic barrier for adsorption. It has to be noted here that the sticking coefficient of adsorbing H$_2$ depends on the orientation as well as the impinging angle of the molecule.\cite{WinklerRendulic,PhysRevB.32.5032} This dependency is not included in the used model. Nevertheless, Eq.~\ref{eq:eqHK} provides an upper limit for the adsorption frequency and hence a lower limit of the time to achieve thermodynamic equilibrium. 
Similarly the desorption frequency reads:
\begin{equation}\label{eq:eqDes}
  \nu_\mathrm{des}\left( T \right) = \nu_0 \cdot \exp{\left(-\frac{E_\mathrm{des}}{k_B T}\right)},
\end{equation}
where the attempt frequency $\nu_0$ is calculated within the harmonic transition state theory:\cite{Vineyard1957}
\begin{equation}\label{eq:eqHTST}
  \nu_0=\frac{\prod_{i=1}^{3N}\nu_i^\mathrm{min}}{\prod_{i=1}^{3N-1}\nu_i^\mathrm{sad}}.
\end{equation}
Here $\nu_i^\mathrm{min}$ are the 3$N$ eigenfrequencies at the minimum and $\nu_i^\mathrm{sad}$ are the $3N$--$1$ nonimaginary eigenfrequencies at the transition point. In order to estimate the attempt frequency we have calculated the dynamical matrix of the 4 topmost atomic layers of an 8 layer thick 2\,$\times$\,2 slab using the small displacements method for a H$_2$ molecule (i) adsorbed at the surface and (ii) at the transition point. The attempt frequency calculated from Eq.~\ref{eq:eqHTST} is $\nu_0$\,=\,5.6\,$\times$\,$10^{13}$\,s$^{-1}$.

\begin{figure}[t]
\includegraphics[width=1.0\columnwidth]{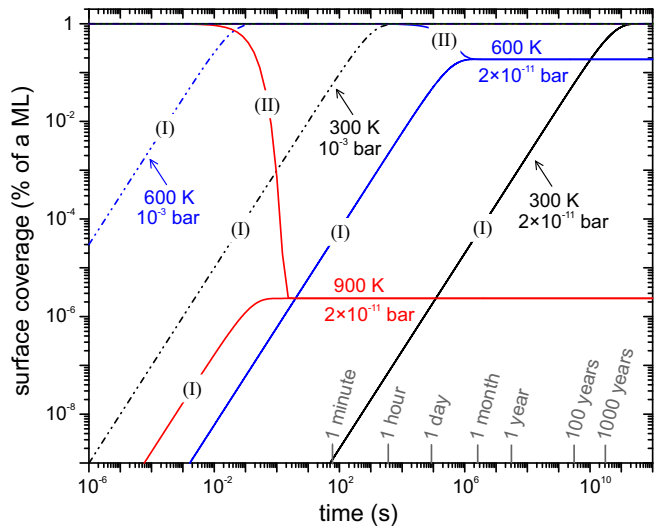}
\caption{\label{fig:fig_coverage} Kinetic surface phase diagram of hydrogen adsorption and desorption on the $m$-plane GaN surface. The hydrogen surface coverage is given as function of time for different H$_2$ pressures and temperatures. (I) and (II) denote clean and fully covered surface initial conditions, respectively.}
\end{figure}

The temporal evolution of the surface coverage is described by the following rate equation:
\begin{equation}\label{eq:eq1}
  \frac{dc}{dt}=(1-c) \cdot \nu_\mathrm{ads}-c \cdot \nu_\mathrm{des},
\end{equation}
where $c$ is the surface coverage and $t$ is the time. Equation (\ref{eq:eq1}) was solved for I $c(t=0)=0$ (clean surface) and for II $c(t=0)=1$ (fully covered surface). The former starting condition (I) corresponds to an adsorption experiment and the latter (II) to a desorption experiment.

In Fig.~\ref{fig:fig_coverage} the coverage is plotted against exposure time to molecular H$_2$ for the two aforementioned initial boundary conditions and for various H$_2$ pressures and temperatures. A striking finding is that although at H$_2$ pressures in the order of 2\,$\times$\,10$^{-11}$\,bar at RT it is thermodynamically favorable to adsorb hydrogen at the surface and the equilibrium coverage is almost 100\%, the time scale to achieve 10\% or 100\% surface coverage is more than a century or a millennium, respectively. Hence, within the time scale of the performed adsorption experiments at RT and partial pressures as low as 10$^{-11}$\,bar (section A) thermodynamic equilibrium between H$_2$ gas and the GaN$(1\overline{1}00)$ surface is kinetically hindered and only atomic hydrogen is able to adsorb quickly and to induce changes in the structural and the electronic surface properties. On the other hand, although the desorption barrier of H$_2$ is larger than the H$_2$ adsorption barrier, desorption of molecular hydrogen can take place in considerably shorter time scales. This can be attributed to (i) the higher temperatures applied to dehydrogenate the surface and (ii) the considerably larger desorption attempt frequency $\nu_0$ than the adsorption attempt frequency $f$ [see Eq.~(\ref{eq:eqHK})]. For example, at the aforementioned temperature and pressure, the adsorption attempt frequency $f$ in Eq.~(\ref{eq:eqHK}) is 2\,$\times$\,10$^{-7}$\,s$^{-1}$ per 1\,$\times$\,1 surface cell area as opposed to the desorption attempt frequency of $\nu_0$\,=\,5.6\,$\times$\,10$^{13}$\,Hz in Eq.~(\ref{eq:eqDes}).

These results are consistent with the observations made in the UHV adsorption/desorption experiments. At low temperatures and low partial pressures thermodynamic equilibrium of the clean $m$-plane GaN surface with a H$_2$ atmosphere is kinetically hindered (compare to the solid black line in Fig.~\ref{fig:fig_coverage} which represents the experimental conditions for the performed experiment of H$_2$ exposure). On the other hand, for the interaction of the clean GaN$(1\overline{1}00)$ surface with activated atomic H thermodynamic equilibrium between the surface and the offered H species can be established within the time scale of a few minutes even at pressures as low as 10$^{-11}$ bar.

\section{SUMMARY}\label{Sum-GaN}

Photoelectron and electron energy loss spectroscopy experiments were combined with first-principles calculations to investigate adsorption and desorption of molecular as well as atomic hydrogen on the nonpolar GaN($1\overline{1}00$) surfaces. Our results show that passivation of the surface cation and anion dangling bonds by hydrogen is thermodynamically favored at room temperature even at hydrogen pressures as low as $10^{-16}$\,bar. Adsorption of molecular hydrogen is associated with a barrier of 0.55\,eV, which leads to unrealistically high exposure times to complete a full monolayer. By contrast, only a few minutes of exposure time are required if the clean $m$-plane GaN surface is brought into an atomic hydrogen atmosphere. On the other hand, hydrogen desorption requires elevated temperatures and is predominantly taking place as hydrogen molecules desorbing from surface dimers. More specifically, our adsorption/desorption experiments confirm that in UHV at room temperature, H-adsorption could only be achieved using atomic hydrogen and desorption starts above $\sim$\,800\,K.
	
The differences in the surface properties of the bare and the H-covered GaN($1\overline{1}00$) surface have been characterized by electron spectroscopy and density functional calculations. Stretching and bending vibrations of the H adatoms at the Ga--N surface dimer structure were identified experimentally and are consistent with the energies and dispersion of calculated surface phonon modes. Furthermore, H adsorption was demonstrated to strongly influence the surface electronic properties. H adatoms occupy the Ga-- and N-- dangling bonds and induce a shift of occupied and unoccupied surface states out of the gap region across the VBM and CBM, respectively, which influences the surface electron depletion layer. Specifically, the raise of unoccupied intragap surface states from below the CBM for the clean surface to energies above the CBM for the H-covered surface induces an unpinning of the surface Fermi level and a reduction of the surface upward band bending from 0.6 to 0.2\,eV. 

These insights demonstrate that chemisorption in a gas exposure experiment or  furthermore the probability of impurity incorporation during crystal growth is strongly dependent on the experimental conditions as well as on the energetics and kinetics of the surface dissociation/adsorption reactions at the surface. For other reactions at GaN surfaces, one might expect comparable kinetically hindered reaction mechanisms that lead to strong deviations between the actually obtained coverage and the supplied reactant load.

\begin{acknowledgments}
This work was supported in part by the Carl Zeiss Stiftung and in part by project ‘PowerBase’. This project has received funding from the Electronic Component Systems for European Leadership Joint Undertaking under grant agreement No 662133. This Joint Undertaking receives support from the European Union’s Horizon 2020 research and innovation programme and Austria, Belgium, Germany, Italy, Netherlands, Norway, Slovakia, Spain and United Kingdom. We thank A.\,M{\"u}ller and T.\,Hannappel for providing access to their vacuum transfer system.
\end{acknowledgments}


\begin{thebibliography}{85}%
\makeatletter
\providecommand \@ifxundefined [1]{%
 \@ifx{#1\undefined}
}%
\providecommand \@ifnum [1]{%
 \ifnum #1\expandafter \@firstoftwo
 \else \expandafter \@secondoftwo
 \fi
}%
\providecommand \@ifx [1]{%
 \ifx #1\expandafter \@firstoftwo
 \else \expandafter \@secondoftwo
 \fi
}%
\providecommand \natexlab [1]{#1}%
\providecommand \enquote  [1]{``#1''}%
\providecommand \bibnamefont  [1]{#1}%
\providecommand \bibfnamefont [1]{#1}%
\providecommand \citenamefont [1]{#1}%
\providecommand \href@noop [0]{\@secondoftwo}%
\providecommand \href [0]{\begingroup \@sanitize@url \@href}%
\providecommand \@href[1]{\@@startlink{#1}\@@href}%
\providecommand \@@href[1]{\endgroup#1\@@endlink}%
\providecommand \@sanitize@url [0]{\catcode `\\12\catcode `\$12\catcode
  `\&12\catcode `\#12\catcode `\^12\catcode `\_12\catcode `\%12\relax}%
\providecommand \@@startlink[1]{}%
\providecommand \@@endlink[0]{}%
\providecommand \url  [0]{\begingroup\@sanitize@url \@url }%
\providecommand \@url [1]{\endgroup\@href {#1}{\urlprefix }}%
\providecommand \urlprefix  [0]{URL }%
\providecommand \Eprint [0]{\href }%
\providecommand \doibase [0]{http://dx.doi.org/}%
\providecommand \selectlanguage [0]{\@gobble}%
\providecommand \bibinfo  [0]{\@secondoftwo}%
\providecommand \bibfield  [0]{\@secondoftwo}%
\providecommand \translation [1]{[#1]}%
\providecommand \BibitemOpen [0]{}%
\providecommand \bibitemStop [0]{}%
\providecommand \bibitemNoStop [0]{.\EOS\space}%
\providecommand \EOS [0]{\spacefactor3000\relax}%
\providecommand \BibitemShut  [1]{\csname bibitem#1\endcsname}%
\let\auto@bib@innerbib\@empty
\bibitem [{\citenamefont {Pimputkar}\ \emph {et~al.}(2009)\citenamefont
  {Pimputkar}, \citenamefont {Speck}, \citenamefont {DenBaars},\ and\
  \citenamefont {Nakamura}}]{Pimputkar2009}%
  \BibitemOpen
  \bibfield  {author} {\bibinfo {author} {\bibfnamefont {S.}~\bibnamefont
  {Pimputkar}}, \bibinfo {author} {\bibfnamefont {J.~S.}\ \bibnamefont
  {Speck}}, \bibinfo {author} {\bibfnamefont {S.~P.}\ \bibnamefont {DenBaars}},
  \ and\ \bibinfo {author} {\bibfnamefont {S.}~\bibnamefont {Nakamura}},\
  }\href@noop {} {\bibfield  {journal} {\bibinfo  {journal} {Nat. Photonics}\
  }\textbf {\bibinfo {volume} {3}},\ \bibinfo {pages} {179} (\bibinfo {year}
  {2009})}\BibitemShut {NoStop}%
\bibitem [{\citenamefont {Crawford}(2009)}]{Crawford2009}%
  \BibitemOpen
  \bibfield  {author} {\bibinfo {author} {\bibfnamefont {M.~H.}\ \bibnamefont
  {Crawford}},\ }\href@noop {} {\bibfield  {journal} {\bibinfo  {journal} {IEEE
  J. Sel. Top. Quant.}\ }\textbf {\bibinfo {volume} {15}},\ \bibinfo {pages}
  {1028} (\bibinfo {year} {2009})}\BibitemShut {NoStop}%
\bibitem [{\citenamefont {Chang}\ \emph {et~al.}(2012)\citenamefont {Chang},
  \citenamefont {Das}, \citenamefont {Varde},\ and\ \citenamefont
  {Pecht}}]{Chang2012}%
  \BibitemOpen
  \bibfield  {author} {\bibinfo {author} {\bibfnamefont {M.-H.}\ \bibnamefont
  {Chang}}, \bibinfo {author} {\bibfnamefont {D.}~\bibnamefont {Das}}, \bibinfo
  {author} {\bibfnamefont {P.V~.}\ \bibnamefont {Varde}}, \ and\ \bibinfo
  {author} {\bibfnamefont {M.}~\bibnamefont {Pecht}},\ }\href@noop {}
  {\bibfield  {journal} {\bibinfo  {journal} {Microelectron. Reliab.}\ }\textbf
  {\bibinfo {volume} {52}},\ \bibinfo {pages} {762} (\bibinfo {year}
  {2012})}\BibitemShut {NoStop}%
\bibitem [{\citenamefont {Mishra}\ \emph {et~al.}(2008)\citenamefont {Mishra},
  \citenamefont {Shen}, \citenamefont {Kazior},\ and\ \citenamefont
  {Wu}}]{Mishra2008}%
  \BibitemOpen
  \bibfield  {author} {\bibinfo {author} {\bibfnamefont {U.~K.}\ \bibnamefont
  {Mishra}}, \bibinfo {author} {\bibfnamefont {L.}~\bibnamefont {Shen}},
  \bibinfo {author} {\bibfnamefont {T.~E.}\ \bibnamefont {Kazior}}, \ and\
  \bibinfo {author} {\bibfnamefont {Y.-F.}\ \bibnamefont {Wu}},\ }\href@noop {}
  {\bibfield  {journal} {\bibinfo  {journal} {IEEE Proc.}\ }\textbf {\bibinfo
  {volume} {96}},\ \bibinfo {pages} {287} (\bibinfo {year} {2008})}\BibitemShut
  {NoStop}%
\bibitem [{\citenamefont {Ikeda}\ \emph {et~al.}(2010)\citenamefont {Ikeda},
  \citenamefont {Niiyama}, \citenamefont {Kambayashi}, \citenamefont {Sato},
  \citenamefont {Nomura}, \citenamefont {Kato},\ and\ \citenamefont
  {Yoshida}}]{Ikeda2010}%
  \BibitemOpen
  \bibfield  {author} {\bibinfo {author} {\bibfnamefont {N.}~\bibnamefont
  {Ikeda}}, \bibinfo {author} {\bibfnamefont {Y.}~\bibnamefont {Niiyama}},
  \bibinfo {author} {\bibfnamefont {H.}~\bibnamefont {Kambayashi}}, \bibinfo
  {author} {\bibfnamefont {Y.}~\bibnamefont {Sato}}, \bibinfo {author}
  {\bibfnamefont {T.}~\bibnamefont {Nomura}}, \bibinfo {author} {\bibfnamefont
  {S.}~\bibnamefont {Kato}}, \ and\ \bibinfo {author} {\bibfnamefont
  {S.}~\bibnamefont {Yoshida}},\ }\href@noop {} {\bibfield  {journal} {\bibinfo
   {journal} {IEEE Proc.}\ }\textbf {\bibinfo {volume} {98}},\ \bibinfo {pages}
  {1151} (\bibinfo {year} {2010})}\BibitemShut {NoStop}%
\bibitem [{\citenamefont {Chowdhury}(2015)}]{Chowdhury2015}%
  \BibitemOpen
  \bibfield  {author} {\bibinfo {author} {\bibfnamefont {S.}~\bibnamefont
  {Chowdhury}},\ }\href@noop {} {\bibfield  {journal} {\bibinfo  {journal}
  {Phys. Status Solidi A}\ }\textbf {\bibinfo {volume} {212}},\ \bibinfo
  {pages} {1066} (\bibinfo {year} {2015})}\BibitemShut {NoStop}%
\bibitem [{\citenamefont {Steinhoff}\ \emph {et~al.}(2003)\citenamefont
  {Steinhoff}, \citenamefont {Hermann}, \citenamefont {Schaff}, \citenamefont
  {Eastman}, \citenamefont {Stutzmann},\ and\ \citenamefont
  {Eickhoff}}]{steinhoff03}%
  \BibitemOpen
  \bibfield  {author} {\bibinfo {author} {\bibfnamefont {G.}~\bibnamefont
  {Steinhoff}}, \bibinfo {author} {\bibfnamefont {M.}~\bibnamefont {Hermann}},
  \bibinfo {author} {\bibfnamefont {W.~J.}\ \bibnamefont {Schaff}}, \bibinfo
  {author} {\bibfnamefont {L.~F.}\ \bibnamefont {Eastman}}, \bibinfo {author}
  {\bibfnamefont {M.}~\bibnamefont {Stutzmann}}, \ and\ \bibinfo {author}
  {\bibfnamefont {M.}~\bibnamefont {Eickhoff}},\ }\href@noop {} {\bibfield
  {journal} {\bibinfo  {journal} {Appl. Phys. Lett.}\ }\textbf {\bibinfo
  {volume} {83}},\ \bibinfo {pages} {177} (\bibinfo {year} {2003})}\BibitemShut
  {NoStop}%
\bibitem [{\citenamefont {Pearton}\ \emph {et~al.}(2004)\citenamefont
  {Pearton}, \citenamefont {Kang}, \citenamefont {Kim}, \citenamefont {Ren},
  \citenamefont {Gila}, \citenamefont {Abernathy}, \citenamefont {Lin},\ and\
  \citenamefont {Chu}}]{pear04}%
  \BibitemOpen
  \bibfield  {author} {\bibinfo {author} {\bibfnamefont {S.~J.}\ \bibnamefont
  {Pearton}}, \bibinfo {author} {\bibfnamefont {B.~S.}\ \bibnamefont {Kang}},
  \bibinfo {author} {\bibfnamefont {S.}~\bibnamefont {Kim}}, \bibinfo {author}
  {\bibfnamefont {F.}~\bibnamefont {Ren}}, \bibinfo {author} {\bibfnamefont
  {B.~P.}\ \bibnamefont {Gila}}, \bibinfo {author} {\bibfnamefont {C.~R.}\
  \bibnamefont {Abernathy}}, \bibinfo {author} {\bibfnamefont {J.}~\bibnamefont
  {Lin}}, \ and\ \bibinfo {author} {\bibfnamefont {S.~N.~G.}\ \bibnamefont
  {Chu}},\ }\href@noop {} {\bibfield  {journal} {\bibinfo  {journal} {J. Phys.:
  Condens. Matter}\ }\textbf {\bibinfo {volume} {16}},\ \bibinfo {pages} {R961}
  (\bibinfo {year} {2004})}\BibitemShut {NoStop}%
\bibitem [{\citenamefont {Chen}\ \emph {et~al.}(2009)\citenamefont {Chen},
  \citenamefont {Ganguly}, \citenamefont {Wang}, \citenamefont {Hsu},
  \citenamefont {Chattopadhyay}, \citenamefont {Hsu}, \citenamefont {Chang},
  \citenamefont {Chen},\ and\ \citenamefont {Chen}}]{chen2009}%
  \BibitemOpen
  \bibfield  {author} {\bibinfo {author} {\bibfnamefont {C.-P.}\ \bibnamefont
  {Chen}}, \bibinfo {author} {\bibfnamefont {A.}~\bibnamefont {Ganguly}},
  \bibinfo {author} {\bibfnamefont {C.-H.}\ \bibnamefont {Wang}}, \bibinfo
  {author} {\bibfnamefont {C.-W.}\ \bibnamefont {Hsu}}, \bibinfo {author}
  {\bibfnamefont {S.}~\bibnamefont {Chattopadhyay}}, \bibinfo {author}
  {\bibfnamefont {Y.-K.}\ \bibnamefont {Hsu}}, \bibinfo {author} {\bibfnamefont
  {Y.-C.}\ \bibnamefont {Chang}}, \bibinfo {author} {\bibfnamefont {K.-H.}\
  \bibnamefont {Chen}}, \ and\ \bibinfo {author} {\bibfnamefont {L.-C.}\
  \bibnamefont {Chen}},\ }\href@noop {} {\bibfield  {journal} {\bibinfo
  {journal} {Anal. Chem.}\ }\textbf {\bibinfo {volume} {81}},\ \bibinfo {pages}
  {362} (\bibinfo {year} {2009})}\BibitemShut {NoStop}%
\bibitem [{\citenamefont {Ganguly}\ \emph {et~al.}(2009)\citenamefont
  {Ganguly}, \citenamefont {Chen}, \citenamefont {Lai}, \citenamefont {Kuo},
  \citenamefont {Hsu}, \citenamefont {Chen},\ and\ \citenamefont
  {Chen}}]{Ganguly2009}%
  \BibitemOpen
  \bibfield  {author} {\bibinfo {author} {\bibfnamefont {A.}~\bibnamefont
  {Ganguly}}, \bibinfo {author} {\bibfnamefont {C.-P.}\ \bibnamefont {Chen}},
  \bibinfo {author} {\bibfnamefont {Y.-T.}\ \bibnamefont {Lai}}, \bibinfo
  {author} {\bibfnamefont {C.-C.}\ \bibnamefont {Kuo}}, \bibinfo {author}
  {\bibfnamefont {C.-W.}\ \bibnamefont {Hsu}}, \bibinfo {author} {\bibfnamefont
  {K.-H.}\ \bibnamefont {Chen}}, \ and\ \bibinfo {author} {\bibfnamefont
  {L.-C.}\ \bibnamefont {Chen}},\ }\href@noop {} {\bibfield  {journal}
  {\bibinfo  {journal} {J. Mater. Chem.}\ }\textbf {\bibinfo {volume} {19}},\
  \bibinfo {pages} {928} (\bibinfo {year} {2009})}\BibitemShut {NoStop}%
\bibitem [{\citenamefont {Pearton}\ \emph {et~al.}({2010})\citenamefont
  {Pearton}, \citenamefont {Ren}, \citenamefont {Wang}, \citenamefont {Chu},
  \citenamefont {Chen}, \citenamefont {Chang}, \citenamefont {Lim},
  \citenamefont {Lin},\ and\ \citenamefont {Norton}}]{pearton2010}%
  \BibitemOpen
  \bibfield  {author} {\bibinfo {author} {\bibfnamefont {S.~J.}\ \bibnamefont
  {Pearton}}, \bibinfo {author} {\bibfnamefont {F.}~\bibnamefont {Ren}},
  \bibinfo {author} {\bibfnamefont {Y.-L.}\ \bibnamefont {Wang}}, \bibinfo
  {author} {\bibfnamefont {B.~H.}\ \bibnamefont {Chu}}, \bibinfo {author}
  {\bibfnamefont {K.~H.}\ \bibnamefont {Chen}}, \bibinfo {author}
  {\bibfnamefont {C.~Y.}\ \bibnamefont {Chang}}, \bibinfo {author}
  {\bibfnamefont {W.}~\bibnamefont {Lim}}, \bibinfo {author} {\bibfnamefont
  {J.}~\bibnamefont {Lin}}, \ and\ \bibinfo {author} {\bibfnamefont {D.~P.}\
  \bibnamefont {Norton}},\ }\href@noop {} {\bibfield  {journal} {\bibinfo
  {journal} {{Prog. Mat. Sci.}}\ }\textbf {\bibinfo {volume} {{55}}},\ \bibinfo
  {pages} {{1}} (\bibinfo {year} {{2010}})}\BibitemShut {NoStop}%
\bibitem [{\citenamefont {Teubert}\ \emph {et~al.}(2011)\citenamefont
  {Teubert}, \citenamefont {Becker}, \citenamefont {Furtmayr},\ and\
  \citenamefont {Eickhoff}}]{Teubert2011}%
  \BibitemOpen
  \bibfield  {author} {\bibinfo {author} {\bibfnamefont {J.}~\bibnamefont
  {Teubert}}, \bibinfo {author} {\bibfnamefont {P.}~\bibnamefont {Becker}},
  \bibinfo {author} {\bibfnamefont {F.}~\bibnamefont {Furtmayr}}, \ and\
  \bibinfo {author} {\bibfnamefont {M.}~\bibnamefont {Eickhoff}},\ }\href@noop
  {} {\bibfield  {journal} {\bibinfo  {journal} {Nanotechnology}\ }\textbf
  {\bibinfo {volume} {22}},\ \bibinfo {pages} {275505} (\bibinfo {year}
  {2011})}\BibitemShut {NoStop}%
\bibitem [{\citenamefont {Guo}\ \emph {et~al.}(2010)\citenamefont {Guo},
  \citenamefont {Zhang}, \citenamefont {Banerjee},\ and\ \citenamefont
  {Bhattacharya}}]{Guo2010}%
  \BibitemOpen
  \bibfield  {author} {\bibinfo {author} {\bibfnamefont {W.}~\bibnamefont
  {Guo}}, \bibinfo {author} {\bibfnamefont {M.}~\bibnamefont {Zhang}}, \bibinfo
  {author} {\bibfnamefont {A.}~\bibnamefont {Banerjee}}, \ and\ \bibinfo
  {author} {\bibfnamefont {P.}~\bibnamefont {Bhattacharya}},\ }\href@noop {}
  {\bibfield  {journal} {\bibinfo  {journal} {Nano Lett.}\ }\textbf {\bibinfo
  {volume} {10}},\ \bibinfo {pages} {3355} (\bibinfo {year}
  {2010})}\BibitemShut {NoStop}%
\bibitem [{\citenamefont {Li}\ and\ \citenamefont {Waag}(2012)}]{li2012}%
  \BibitemOpen
  \bibfield  {author} {\bibinfo {author} {\bibfnamefont {S.}~\bibnamefont
  {Li}}\ and\ \bibinfo {author} {\bibfnamefont {A.}~\bibnamefont {Waag}},\
  }\href@noop {} {\bibfield  {journal} {\bibinfo  {journal} {J. Appl. Phys.}\
  }\textbf {\bibinfo {volume} {111}},\ \bibinfo {eid} {071101} (\bibinfo {year}
  {2012})}\BibitemShut {NoStop}%
\bibitem [{\citenamefont {Gonz\'{a}lez-Posada}\ \emph
  {et~al.}(2012)\citenamefont {Gonz\'{a}lez-Posada}, \citenamefont {Songmuang},
  \citenamefont {Hertog},\ and\ \citenamefont {Monroy}}]{gonzales2012}%
  \BibitemOpen
  \bibfield  {author} {\bibinfo {author} {\bibfnamefont {F.}~\bibnamefont
  {Gonz\'{a}lez-Posada}}, \bibinfo {author} {\bibfnamefont {R.}~\bibnamefont
  {Songmuang}}, \bibinfo {author} {\bibfnamefont {M.~Den}\ \bibnamefont
  {Hertog}}, \ and\ \bibinfo {author} {\bibfnamefont {E.}~\bibnamefont
  {Monroy}},\ }\href@noop {} {\bibfield  {journal} {\bibinfo  {journal} {Nano
  Lett.}\ }\textbf {\bibinfo {volume} {12}},\ \bibinfo {pages} {172} (\bibinfo
  {year} {2012})}\BibitemShut {NoStop}%
\bibitem [{\citenamefont {Shen}\ \emph {et~al.}(2010)\citenamefont {Shen},
  \citenamefont {Small}, \citenamefont {Wang}, \citenamefont {Allen},
  \citenamefont {Fernandez-Serra}, \citenamefont {Hybertsen},\ and\
  \citenamefont {Muckerman}}]{Shen10JPC}%
  \BibitemOpen
  \bibfield  {author} {\bibinfo {author} {\bibfnamefont {X.}~\bibnamefont
  {Shen}}, \bibinfo {author} {\bibfnamefont {Y.~A.}\ \bibnamefont {Small}},
  \bibinfo {author} {\bibfnamefont {J.}~\bibnamefont {Wang}}, \bibinfo {author}
  {\bibfnamefont {P.~B.}\ \bibnamefont {Allen}}, \bibinfo {author}
  {\bibfnamefont {M.~V.}\ \bibnamefont {Fernandez-Serra}}, \bibinfo {author}
  {\bibfnamefont {M.~S.}\ \bibnamefont {Hybertsen}}, \ and\ \bibinfo {author}
  {\bibfnamefont {J.~T.}\ \bibnamefont {Muckerman}},\ }\href@noop {} {\bibfield
   {journal} {\bibinfo  {journal} {J. Phys. Chem. C}\ }\textbf {\bibinfo
  {volume} {114}},\ \bibinfo {pages} {13695} (\bibinfo {year}
  {2010})}\BibitemShut {NoStop}%
\bibitem [{\citenamefont {Wang}\ \emph {et~al.}(2011)\citenamefont {Wang},
  \citenamefont {Pierre}, \citenamefont {Kibria}, \citenamefont {Cui},
  \citenamefont {Han}, \citenamefont {Bevan}, \citenamefont {Guo},
  \citenamefont {Paradis}, \citenamefont {Hakima},\ and\ \citenamefont
  {Mi}}]{Wang2011}%
  \BibitemOpen
  \bibfield  {author} {\bibinfo {author} {\bibfnamefont {D.}~\bibnamefont
  {Wang}}, \bibinfo {author} {\bibfnamefont {A.}~\bibnamefont {Pierre}},
  \bibinfo {author} {\bibfnamefont {M.~G.}\ \bibnamefont {Kibria}}, \bibinfo
  {author} {\bibfnamefont {K.}~\bibnamefont {Cui}}, \bibinfo {author}
  {\bibfnamefont {X.}~\bibnamefont {Han}}, \bibinfo {author} {\bibfnamefont
  {K.~H.}\ \bibnamefont {Bevan}}, \bibinfo {author} {\bibfnamefont
  {H.}~\bibnamefont {Guo}}, \bibinfo {author} {\bibfnamefont {S.}~\bibnamefont
  {Paradis}}, \bibinfo {author} {\bibfnamefont {A.-R.}\ \bibnamefont {Hakima}},
  \ and\ \bibinfo {author} {\bibfnamefont {Z.}~\bibnamefont {Mi}},\ }\href@noop
  {} {\bibfield  {journal} {\bibinfo  {journal} {Nano Lett.}\ }\textbf
  {\bibinfo {volume} {11}},\ \bibinfo {pages} {2353} (\bibinfo {year}
  {2011})}\BibitemShut {NoStop}%
\bibitem [{\citenamefont {Wang}\ \emph {et~al.}(2012)\citenamefont {Wang},
  \citenamefont {Pedroza}, \citenamefont {Poissier},\ and\ \citenamefont
  {Fern\'{a}ndez-Serra}}]{Wang12JPC}%
  \BibitemOpen
  \bibfield  {author} {\bibinfo {author} {\bibfnamefont {J.}~\bibnamefont
  {Wang}}, \bibinfo {author} {\bibfnamefont {L.S.}\ \bibnamefont {Pedroza}},
  \bibinfo {author} {\bibfnamefont {A.}~\bibnamefont {Poissier}}, \ and\
  \bibinfo {author} {\bibfnamefont {M.~V.}\ \bibnamefont
  {Fern\'{a}ndez-Serra}},\ }\href@noop {} {\bibfield  {journal} {\bibinfo
  {journal} {J. Phys. Chem. C}\ }\textbf {\bibinfo {volume} {116}},\ \bibinfo
  {pages} {14382} (\bibinfo {year} {2012})}\BibitemShut {NoStop}%
\bibitem [{\citenamefont {Tang}\ \emph {et~al.}(2008)\citenamefont {Tang},
  \citenamefont {Chen}, \citenamefont {Song}, \citenamefont {Lee},
  \citenamefont {Cong}, \citenamefont {Cheng}, \citenamefont {Zhang},
  \citenamefont {Bello},\ and\ \citenamefont {Lee}}]{Tang2008}%
  \BibitemOpen
  \bibfield  {author} {\bibinfo {author} {\bibfnamefont {Y.~B.}\ \bibnamefont
  {Tang}}, \bibinfo {author} {\bibfnamefont {Z.~H.}\ \bibnamefont {Chen}},
  \bibinfo {author} {\bibfnamefont {H.~S.}\ \bibnamefont {Song}}, \bibinfo
  {author} {\bibfnamefont {C.~S.}\ \bibnamefont {Lee}}, \bibinfo {author}
  {\bibfnamefont {H.~T.}\ \bibnamefont {Cong}}, \bibinfo {author}
  {\bibfnamefont {H.~M.}\ \bibnamefont {Cheng}}, \bibinfo {author}
  {\bibfnamefont {W.~J.}\ \bibnamefont {Zhang}}, \bibinfo {author}
  {\bibfnamefont {I.}~\bibnamefont {Bello}}, \ and\ \bibinfo {author}
  {\bibfnamefont {S.~T.}\ \bibnamefont {Lee}},\ }\href@noop {} {\bibfield
  {journal} {\bibinfo  {journal} {Nano Lett.}\ }\textbf {\bibinfo {volume}
  {8}},\ \bibinfo {pages} {4191} (\bibinfo {year} {2008})}\BibitemShut
  {NoStop}%
\bibitem [{\citenamefont {Dong}\ \emph {et~al.}(2009)\citenamefont {Dong},
  \citenamefont {Tian}, \citenamefont {Kempa},\ and\ \citenamefont
  {Lieber}}]{Dong2009}%
  \BibitemOpen
  \bibfield  {author} {\bibinfo {author} {\bibfnamefont {Y.}~\bibnamefont
  {Dong}}, \bibinfo {author} {\bibfnamefont {B.}~\bibnamefont {Tian}}, \bibinfo
  {author} {\bibfnamefont {T.~J.}\ \bibnamefont {Kempa}}, \ and\ \bibinfo
  {author} {\bibfnamefont {C.~M.}\ \bibnamefont {Lieber}},\ }\href@noop {}
  {\bibfield  {journal} {\bibinfo  {journal} {Nano Lett.}\ }\textbf {\bibinfo
  {volume} {9}},\ \bibinfo {pages} {2183} (\bibinfo {year} {2009})}\BibitemShut
  {NoStop}%
\bibitem [{\citenamefont {Wang}\ \emph {et~al.}(2006)\citenamefont {Wang},
  \citenamefont {Talin}, \citenamefont {Werder}, \citenamefont {Creighton},
  \citenamefont {Lai}, \citenamefont {Anderson},\ and\ \citenamefont
  {Arslan}}]{Wang2006Nanotech}%
  \BibitemOpen
  \bibfield  {author} {\bibinfo {author} {\bibfnamefont {G.~T.}\ \bibnamefont
  {Wang}}, \bibinfo {author} {\bibfnamefont {A.~A.}\ \bibnamefont {Talin}},
  \bibinfo {author} {\bibfnamefont {D.~J.}\ \bibnamefont {Werder}}, \bibinfo
  {author} {\bibfnamefont {J.~R.}\ \bibnamefont {Creighton}}, \bibinfo {author}
  {\bibfnamefont {E.}~\bibnamefont {Lai}}, \bibinfo {author} {\bibfnamefont
  {R.~J.}\ \bibnamefont {Anderson}}, \ and\ \bibinfo {author} {\bibfnamefont
  {I.}~\bibnamefont {Arslan}},\ }\href@noop {} {\bibfield  {journal} {\bibinfo
  {journal} {Nanotechnology}\ }\textbf {\bibinfo {volume} {17}},\ \bibinfo
  {pages} {5773} (\bibinfo {year} {2006})}\BibitemShut {NoStop}%
\bibitem [{\citenamefont {Songmuang}\ \emph {et~al.}(2007)\citenamefont
  {Songmuang}, \citenamefont {Landr{\'e}},\ and\ \citenamefont
  {Daudin}}]{Songmuang2007}%
  \BibitemOpen
  \bibfield  {author} {\bibinfo {author} {\bibfnamefont {R.}~\bibnamefont
  {Songmuang}}, \bibinfo {author} {\bibfnamefont {O.}~\bibnamefont
  {Landr{\'e}}}, \ and\ \bibinfo {author} {\bibfnamefont {B.}~\bibnamefont
  {Daudin}},\ }\href@noop {} {\bibfield  {journal} {\bibinfo  {journal} {Appl.
  Phys. Lett.}\ }\textbf {\bibinfo {volume} {91}},\ \bibinfo {eid} {251902}
  (\bibinfo {year} {2007})}\BibitemShut {NoStop}%
\bibitem [{\citenamefont {Schlager}\ \emph {et~al.}(2008)\citenamefont
  {Schlager}, \citenamefont {Bertness}, \citenamefont {Blanchard},
  \citenamefont {Robins}, \citenamefont {Roshko},\ and\ \citenamefont
  {Sanford}}]{Schlager2008}%
  \BibitemOpen
  \bibfield  {author} {\bibinfo {author} {\bibfnamefont {J.~B.}\ \bibnamefont
  {Schlager}}, \bibinfo {author} {\bibfnamefont {K.~A.}\ \bibnamefont
  {Bertness}}, \bibinfo {author} {\bibfnamefont {P.~T.}\ \bibnamefont
  {Blanchard}}, \bibinfo {author} {\bibfnamefont {L.~H.}\ \bibnamefont
  {Robins}}, \bibinfo {author} {\bibfnamefont {A.}~\bibnamefont {Roshko}}, \
  and\ \bibinfo {author} {\bibfnamefont {N.~A.}\ \bibnamefont {Sanford}},\
  }\href@noop {} {\bibfield  {journal} {\bibinfo  {journal} {J. Appl. Phys.}\
  }\textbf {\bibinfo {volume} {103}},\ \bibinfo {eid} {124309} (\bibinfo {year}
  {2008})}\BibitemShut {NoStop}%
\bibitem [{\citenamefont {Ch\`{e}ze}\ \emph {et~al.}(2010)\citenamefont
  {Ch\`{e}ze}, \citenamefont {Geelhaar}, \citenamefont {Brandt}, \citenamefont
  {Weber}, \citenamefont {Riechert}, \citenamefont {M{\"u}nch}, \citenamefont
  {Rothemund}, \citenamefont {Reitzenstein}, \citenamefont {Forchel},
  \citenamefont {Kehagias}, \citenamefont {Komninou}, \citenamefont
  {Dimitrakopulos},\ and\ \citenamefont {Karakostas}}]{cheze2010}%
  \BibitemOpen
  \bibfield  {author} {\bibinfo {author} {\bibfnamefont {C.}~\bibnamefont
  {Ch\`{e}ze}}, \bibinfo {author} {\bibfnamefont {L.}~\bibnamefont {Geelhaar}},
  \bibinfo {author} {\bibfnamefont {O.}~\bibnamefont {Brandt}}, \bibinfo
  {author} {\bibfnamefont {W.~M.}\ \bibnamefont {Weber}}, \bibinfo {author}
  {\bibfnamefont {H.}~\bibnamefont {Riechert}}, \bibinfo {author}
  {\bibfnamefont {S.}~\bibnamefont {M{\"u}nch}}, \bibinfo {author}
  {\bibfnamefont {R.}~\bibnamefont {Rothemund}}, \bibinfo {author}
  {\bibfnamefont {S.}~\bibnamefont {Reitzenstein}}, \bibinfo {author}
  {\bibfnamefont {A.}~\bibnamefont {Forchel}}, \bibinfo {author} {\bibfnamefont
  {T.}~\bibnamefont {Kehagias}}, \bibinfo {author} {\bibfnamefont
  {P.}~\bibnamefont {Komninou}}, \bibinfo {author} {\bibfnamefont {G.~P.}\
  \bibnamefont {Dimitrakopulos}}, \ and\ \bibinfo {author} {\bibfnamefont
  {T.}~\bibnamefont {Karakostas}},\ }\href@noop {} {\bibfield  {journal}
  {\bibinfo  {journal} {Nano Res.}\ }\textbf {\bibinfo {volume} {3}},\ \bibinfo
  {pages} {528} (\bibinfo {year} {2010})}\BibitemShut {NoStop}%
\bibitem [{\citenamefont {Qian}\ \emph {et~al.}(2008)\citenamefont {Qian},
  \citenamefont {Li}, \citenamefont {Gradecak}, \citenamefont {Park},
  \citenamefont {Dong}, \citenamefont {Ding}, \citenamefont {Wang},\ and\
  \citenamefont {Lieber}}]{Qian2008}%
  \BibitemOpen
  \bibfield  {author} {\bibinfo {author} {\bibfnamefont {F.}~\bibnamefont
  {Qian}}, \bibinfo {author} {\bibfnamefont {Y.}~\bibnamefont {Li}}, \bibinfo
  {author} {\bibfnamefont {S.}~\bibnamefont {Gradecak}}, \bibinfo {author}
  {\bibfnamefont {H.-G.}\ \bibnamefont {Park}}, \bibinfo {author}
  {\bibfnamefont {Y.}~\bibnamefont {Dong}}, \bibinfo {author} {\bibfnamefont
  {Y.}~\bibnamefont {Ding}}, \bibinfo {author} {\bibfnamefont {Z.~L.}\
  \bibnamefont {Wang}}, \ and\ \bibinfo {author} {\bibfnamefont {C.~M.}\
  \bibnamefont {Lieber}},\ }\href@noop {} {\bibfield  {journal} {\bibinfo
  {journal} {Nat. Mater.}\ }\textbf {\bibinfo {volume} {7}},\ \bibinfo {pages}
  {701} (\bibinfo {year} {2008})}\BibitemShut {NoStop}%
\bibitem [{\citenamefont {Koester}\ \emph {et~al.}(2011)\citenamefont
  {Koester}, \citenamefont {Hwang}, \citenamefont {Salomon}, \citenamefont
  {Chen}, \citenamefont {Bougerol}, \citenamefont {Barnes}, \citenamefont
  {Dang}, \citenamefont {Rigutti}, \citenamefont {de~Luna~Bugallo},
  \citenamefont {Jacopin}, \citenamefont {Tchernycheva}, \citenamefont
  {Durand},\ and\ \citenamefont {Eymery}}]{Koester2011}%
  \BibitemOpen
  \bibfield  {author} {\bibinfo {author} {\bibfnamefont {R.}~\bibnamefont
  {Koester}}, \bibinfo {author} {\bibfnamefont {J.-S.}\ \bibnamefont {Hwang}},
  \bibinfo {author} {\bibfnamefont {D.}~\bibnamefont {Salomon}}, \bibinfo
  {author} {\bibfnamefont {X.}~\bibnamefont {Chen}}, \bibinfo {author}
  {\bibfnamefont {C.}~\bibnamefont {Bougerol}}, \bibinfo {author}
  {\bibfnamefont {J.-P.}\ \bibnamefont {Barnes}}, \bibinfo {author}
  {\bibfnamefont {D.~L.}\ \bibnamefont {Dang}}, \bibinfo {author}
  {\bibfnamefont {L.}~\bibnamefont {Rigutti}}, \bibinfo {author} {\bibfnamefont
  {A.}~\bibnamefont {de~Luna~Bugallo}}, \bibinfo {author} {\bibfnamefont
  {G.}~\bibnamefont {Jacopin}}, \bibinfo {author} {\bibfnamefont
  {M.}~\bibnamefont {Tchernycheva}}, \bibinfo {author} {\bibfnamefont
  {C.}~\bibnamefont {Durand}}, \ and\ \bibinfo {author} {\bibfnamefont
  {J.}~\bibnamefont {Eymery}},\ }\href@noop {} {\bibfield  {journal} {\bibinfo
  {journal} {Nano Lett.}\ }\textbf {\bibinfo {volume} {11}},\ \bibinfo {pages}
  {4839} (\bibinfo {year} {2011})}\BibitemShut {NoStop}%
\bibitem [{\citenamefont {Yeh}\ \emph {et~al.}(2012)\citenamefont {Yeh},
  \citenamefont {Lin}, \citenamefont {Stewart}, \citenamefont {Dapkus},
  \citenamefont {Sarkissian}, \citenamefont {O'Brien}, \citenamefont {Ahn},\
  and\ \citenamefont {Nutt}}]{Yeh2012}%
  \BibitemOpen
  \bibfield  {author} {\bibinfo {author} {\bibfnamefont {T.-W.}\ \bibnamefont
  {Yeh}}, \bibinfo {author} {\bibfnamefont {Y.-T.}\ \bibnamefont {Lin}},
  \bibinfo {author} {\bibfnamefont {L.~S.}\ \bibnamefont {Stewart}}, \bibinfo
  {author} {\bibfnamefont {P.~D.}\ \bibnamefont {Dapkus}}, \bibinfo {author}
  {\bibfnamefont {R.}~\bibnamefont {Sarkissian}}, \bibinfo {author}
  {\bibfnamefont {J.~D.}\ \bibnamefont {O'Brien}}, \bibinfo {author}
  {\bibfnamefont {B.}~\bibnamefont {Ahn}}, \ and\ \bibinfo {author}
  {\bibfnamefont {S.~R.}\ \bibnamefont {Nutt}},\ }\href@noop {} {\bibfield
  {journal} {\bibinfo  {journal} {Nano Lett.}\ }\textbf {\bibinfo {volume}
  {12}},\ \bibinfo {pages} {3257} (\bibinfo {year} {2012})}\BibitemShut
  {NoStop}%
\bibitem [{\citenamefont {Nguyen}\ \emph {et~al.}(2011)\citenamefont {Nguyen},
  \citenamefont {Zhang}, \citenamefont {Cui}, \citenamefont {Han},
  \citenamefont {Fathololoumi}, \citenamefont {Couillard}, \citenamefont
  {Botton},\ and\ \citenamefont {Mi}}]{Nguyen2011}%
  \BibitemOpen
  \bibfield  {author} {\bibinfo {author} {\bibfnamefont {H.~P.~T.}\
  \bibnamefont {Nguyen}}, \bibinfo {author} {\bibfnamefont {S.}~\bibnamefont
  {Zhang}}, \bibinfo {author} {\bibfnamefont {K.}~\bibnamefont {Cui}}, \bibinfo
  {author} {\bibfnamefont {X.}~\bibnamefont {Han}}, \bibinfo {author}
  {\bibfnamefont {S.}~\bibnamefont {Fathololoumi}}, \bibinfo {author}
  {\bibfnamefont {M.}~\bibnamefont {Couillard}}, \bibinfo {author}
  {\bibfnamefont {G.~A.}\ \bibnamefont {Botton}}, \ and\ \bibinfo {author}
  {\bibfnamefont {Z.}~\bibnamefont {Mi}},\ }\href@noop {} {\bibfield  {journal}
  {\bibinfo  {journal} {Nano Lett.}\ }\textbf {\bibinfo {volume} {11}},\
  \bibinfo {pages} {1919} (\bibinfo {year} {2011})}\BibitemShut {NoStop}%
\bibitem [{\citenamefont {Kehagias}\ \emph {et~al.}(2013)\citenamefont
  {Kehagias}, \citenamefont {Dimitrakopulos}, \citenamefont {Becker},
  \citenamefont {Kioseoglou}, \citenamefont {Furtmayr}, \citenamefont
  {Koukoula}, \citenamefont {H{\"a}usler}, \citenamefont {Chernikov},
  \citenamefont {Chatterjee}, \citenamefont {Karakostas}, \citenamefont
  {Solowan}, \citenamefont {Schwarz}, \citenamefont {Eickhoff},\ and\
  \citenamefont {Komninou}}]{Kehagias2013}%
  \BibitemOpen
  \bibfield  {author} {\bibinfo {author} {\bibfnamefont {T.}~\bibnamefont
  {Kehagias}}, \bibinfo {author} {\bibfnamefont {G.~P.}\ \bibnamefont
  {Dimitrakopulos}}, \bibinfo {author} {\bibfnamefont {P.}~\bibnamefont
  {Becker}}, \bibinfo {author} {\bibfnamefont {J.}~\bibnamefont {Kioseoglou}},
  \bibinfo {author} {\bibfnamefont {F.}~\bibnamefont {Furtmayr}}, \bibinfo
  {author} {\bibfnamefont {T.}~\bibnamefont {Koukoula}}, \bibinfo {author}
  {\bibfnamefont {I.}~\bibnamefont {H{\"a}usler}}, \bibinfo {author}
  {\bibfnamefont {A.}~\bibnamefont {Chernikov}}, \bibinfo {author}
  {\bibfnamefont {S.}~\bibnamefont {Chatterjee}}, \bibinfo {author}
  {\bibfnamefont {T.}~\bibnamefont {Karakostas}}, \bibinfo {author}
  {\bibfnamefont {H.-M.}\ \bibnamefont {Solowan}}, \bibinfo {author}
  {\bibfnamefont {U.~T.}\ \bibnamefont {Schwarz}}, \bibinfo {author}
  {\bibfnamefont {M.}~\bibnamefont {Eickhoff}}, \ and\ \bibinfo {author}
  {\bibfnamefont {P.}~\bibnamefont {Komninou}},\ }\href@noop {} {\bibfield
  {journal} {\bibinfo  {journal} {Nanotechnology}\ }\textbf {\bibinfo {volume}
  {24}},\ \bibinfo {pages} {435702} (\bibinfo {year} {2013})}\BibitemShut
  {NoStop}%
\bibitem [{\citenamefont {Sch{\"o}rmann}\ \emph {et~al.}(2013)\citenamefont
  {Sch{\"o}rmann}, \citenamefont {Hille}, \citenamefont {Sch{\"a}fer},
  \citenamefont {M{\"u}��ener}, \citenamefont {Becker}, \citenamefont
  {Klar}, \citenamefont {Kleine-Boymann}, \citenamefont {Rohnke}, \citenamefont
  {de~la Mata}, \citenamefont {Arbiol}, \citenamefont {Hofmann}, \citenamefont
  {Teubert},\ and\ \citenamefont {Eickhoff}}]{Schoermann2013}%
  \BibitemOpen
  \bibfield  {author} {\bibinfo {author} {\bibfnamefont {J.}~\bibnamefont
  {Sch{\"o}rmann}}, \bibinfo {author} {\bibfnamefont {P.}~\bibnamefont
  {Hille}}, \bibinfo {author} {\bibfnamefont {M.}~\bibnamefont {Sch{\"a}fer}},
  \bibinfo {author} {\bibfnamefont {J.}~\bibnamefont {M{\"u}��ener}},
  \bibinfo {author} {\bibfnamefont {P.}~\bibnamefont {Becker}}, \bibinfo
  {author} {\bibfnamefont {P.~J.}\ \bibnamefont {Klar}}, \bibinfo {author}
  {\bibfnamefont {M.}~\bibnamefont {Kleine-Boymann}}, \bibinfo {author}
  {\bibfnamefont {M.}~\bibnamefont {Rohnke}}, \bibinfo {author} {\bibfnamefont
  {M.}~\bibnamefont {de~la Mata}}, \bibinfo {author} {\bibfnamefont
  {J.}~\bibnamefont {Arbiol}}, \bibinfo {author} {\bibfnamefont {D.~M.}\
  \bibnamefont {Hofmann}}, \bibinfo {author} {\bibfnamefont {J.}~\bibnamefont
  {Teubert}}, \ and\ \bibinfo {author} {\bibfnamefont {M.}~\bibnamefont
  {Eickhoff}},\ }\href@noop {} {\bibfield  {journal} {\bibinfo  {journal} {J.
  Appl. Phys.}\ }\textbf {\bibinfo {volume} {114}},\ \bibinfo {eid} {103505}
  (\bibinfo {year} {2013})}\BibitemShut {NoStop}%
\bibitem [{\citenamefont {Eller}\ \emph {et~al.}(2013)\citenamefont {Eller},
  \citenamefont {Yang},\ and\ \citenamefont {Nemanich}}]{Eller2013}%
  \BibitemOpen
  \bibfield  {author} {\bibinfo {author} {\bibfnamefont {B.~S.}\ \bibnamefont
  {Eller}}, \bibinfo {author} {\bibfnamefont {J.}~\bibnamefont {Yang}}, \ and\
  \bibinfo {author} {\bibfnamefont {R.~J.}\ \bibnamefont {Nemanich}},\
  }\href@noop {} {\bibfield  {journal} {\bibinfo  {journal} {J. Vac. Sci.
  Technol. A}\ }\textbf {\bibinfo {volume} {31}},\ \bibinfo {pages} {050807}
  (\bibinfo {year} {2013})}\BibitemShut {NoStop}%
\bibitem [{\citenamefont {Zhang}\ and\ \citenamefont {Jr.}(2012)}]{Zhang2012}%
  \BibitemOpen
  \bibfield  {author} {\bibinfo {author} {\bibfnamefont {Z.}~\bibnamefont
  {Zhang}}\ and\ \bibinfo {author} {\bibfnamefont {J.~T.~Yates}\ \bibnamefont
  {Jr.}},\ }\href@noop {} {\bibfield  {journal} {\bibinfo  {journal} {Chem.
  Rev.}\ }\textbf {\bibinfo {volume} {112}},\ \bibinfo {pages} {5520} (\bibinfo
  {year} {2012})}\BibitemShut {NoStop}%
\bibitem [{\citenamefont {Robertson}(2013)}]{Robertson2013}%
  \BibitemOpen
  \bibfield  {author} {\bibinfo {author} {\bibfnamefont {J.}~\bibnamefont
  {Robertson}},\ }\href@noop {} {\bibfield  {journal} {\bibinfo  {journal} {J.
  Vac. Sci. Technol. A}\ }\textbf {\bibinfo {volume} {31}},\ \bibinfo {eid}
  {050821} (\bibinfo {year} {2013})}\BibitemShut {NoStop}%
\bibitem [{\citenamefont {Sanford}\ \emph {et~al.}(2013)\citenamefont
  {Sanford}, \citenamefont {Robins}, \citenamefont {Blanchard}, \citenamefont
  {Soria}, \citenamefont {Klein}, \citenamefont {Eller}, \citenamefont
  {Bertness}, \citenamefont {Schlager},\ and\ \citenamefont
  {Sanders}}]{san2013}%
  \BibitemOpen
  \bibfield  {author} {\bibinfo {author} {\bibfnamefont {N.~A.}\ \bibnamefont
  {Sanford}}, \bibinfo {author} {\bibfnamefont {L.~H.}\ \bibnamefont {Robins}},
  \bibinfo {author} {\bibfnamefont {P.~T.}\ \bibnamefont {Blanchard}}, \bibinfo
  {author} {\bibfnamefont {K.}~\bibnamefont {Soria}}, \bibinfo {author}
  {\bibfnamefont {B.}~\bibnamefont {Klein}}, \bibinfo {author} {\bibfnamefont
  {B.~S.}\ \bibnamefont {Eller}}, \bibinfo {author} {\bibfnamefont {K.~A.}\
  \bibnamefont {Bertness}}, \bibinfo {author} {\bibfnamefont {J.~B.}\
  \bibnamefont {Schlager}}, \ and\ \bibinfo {author} {\bibfnamefont {A.~W.}\
  \bibnamefont {Sanders}},\ }\href@noop {} {\bibfield  {journal} {\bibinfo
  {journal} {J. Appl. Phys.}\ }\textbf {\bibinfo {volume} {113}},\ \bibinfo
  {eid} {174306} (\bibinfo {year} {2013})}\BibitemShut {NoStop}%
\bibitem [{\citenamefont {Neugebauer}\ and\ \citenamefont {Van~de
  Walle}(1996)}]{Neugebauer1996}%
  \BibitemOpen
  \bibfield  {author} {\bibinfo {author} {\bibfnamefont {J.}~\bibnamefont
  {Neugebauer}}\ and\ \bibinfo {author} {\bibfnamefont {C.~G.}\ \bibnamefont
  {Van~de Walle}},\ }\href@noop {} {\bibfield  {journal} {\bibinfo  {journal}
  {Appl. Phys. Lett.}\ }\textbf {\bibinfo {volume} {68}},\ \bibinfo {pages}
  {1829} (\bibinfo {year} {1996})}\BibitemShut {NoStop}%
\bibitem [{\citenamefont {Pearton}\ and\ \citenamefont
  {Lee}(1999)}]{Pearton1999}%
  \BibitemOpen
  \bibfield  {author} {\bibinfo {author} {\bibfnamefont {S.~J.}\ \bibnamefont
  {Pearton}}\ and\ \bibinfo {author} {\bibfnamefont {J.~W.}\ \bibnamefont
  {Lee}},\ }\bibfield  {title} {\enquote {\bibinfo {title} {{The Properties of
  Hydrogen in GaN and Related Alloys}},}\ }in\ \href@noop {} {\emph {\bibinfo
  {booktitle} {Hydrogen in Semiconductors II}}},\ \bibinfo {series}
  {Semiconductors and Semimetals}, Vol.~\bibinfo {volume} {61},\ \bibinfo
  {editor} {edited by\ \bibinfo {editor} {\bibfnamefont {N.~H.}\ \bibnamefont
  {Nickel}}}\ (\bibinfo  {publisher} {Elsevier},\ \bibinfo {year} {1999})\ p.\
  \bibinfo {pages} {441}\BibitemShut {NoStop}%
\bibitem [{\citenamefont {Neugebauer}\ and\ \citenamefont
  {de~Walle}(1999)}]{Neugebauer1999}%
  \BibitemOpen
  \bibfield  {author} {\bibinfo {author} {\bibfnamefont {J.}~\bibnamefont
  {Neugebauer}}\ and\ \bibinfo {author} {\bibfnamefont {C.~G.~Van}\
  \bibnamefont {de~Walle}},\ }\bibfield  {title} {\enquote {\bibinfo {title}
  {{Theory of Hydrogen in GaN}},}\ }in\ \href@noop {} {\emph {\bibinfo
  {booktitle} {Hydrogen in Semiconductors II}}},\ \bibinfo {series}
  {Semiconductors and Semimetals}, Vol.~\bibinfo {volume} {61},\ \bibinfo
  {editor} {edited by\ \bibinfo {editor} {\bibfnamefont {N.~H.}\ \bibnamefont
  {Nickel}}}\ (\bibinfo  {publisher} {Elsevier},\ \bibinfo {year} {1999})\ p.\
  \bibinfo {pages} {479}\BibitemShut {NoStop}%
\bibitem [{\citenamefont {Ambacher}\ \emph {et~al.}(1997)\citenamefont
  {Ambacher}, \citenamefont {Angerer}, \citenamefont {Dimitrov}, \citenamefont
  {Rieger}, \citenamefont {Stutzmann}, \citenamefont {Dollinger},\ and\
  \citenamefont {Bergmaier}}]{Ambacher1997}%
  \BibitemOpen
  \bibfield  {author} {\bibinfo {author} {\bibfnamefont {O.}~\bibnamefont
  {Ambacher}}, \bibinfo {author} {\bibfnamefont {H.}~\bibnamefont {Angerer}},
  \bibinfo {author} {\bibfnamefont {R.}~\bibnamefont {Dimitrov}}, \bibinfo
  {author} {\bibfnamefont {W.}~\bibnamefont {Rieger}}, \bibinfo {author}
  {\bibfnamefont {M.}~\bibnamefont {Stutzmann}}, \bibinfo {author}
  {\bibfnamefont {G.}~\bibnamefont {Dollinger}}, \ and\ \bibinfo {author}
  {\bibfnamefont {A.}~\bibnamefont {Bergmaier}},\ }\href@noop {} {\bibfield
  {journal} {\bibinfo  {journal} {Phys. Status Solidi A}\ }\textbf {\bibinfo
  {volume} {159}},\ \bibinfo {pages} {105} (\bibinfo {year}
  {1997})}\BibitemShut {NoStop}%
\bibitem [{\citenamefont {Okamoto}\ \emph {et~al.}(1999)\citenamefont
  {Okamoto}, \citenamefont {Hashiguchi}, \citenamefont {Okada},\ and\
  \citenamefont {Kawabe}}]{Okamoto1999}%
  \BibitemOpen
  \bibfield  {author} {\bibinfo {author} {\bibfnamefont {Y.}~\bibnamefont
  {Okamoto}}, \bibinfo {author} {\bibfnamefont {S.}~\bibnamefont {Hashiguchi}},
  \bibinfo {author} {\bibfnamefont {Y.}~\bibnamefont {Okada}}, \ and\ \bibinfo
  {author} {\bibfnamefont {M.}~\bibnamefont {Kawabe}},\ }\href@noop {}
  {\bibfield  {journal} {\bibinfo  {journal} {Jap. J. Appl. Phys.}\ }\textbf
  {\bibinfo {volume} {38}},\ \bibinfo {pages} {L230} (\bibinfo {year}
  {1999})}\BibitemShut {NoStop}%
\bibitem [{\citenamefont {Aujol}\ \emph {et~al.}(2001)\citenamefont {Aujol},
  \citenamefont {Trassoudaine}, \citenamefont {Siozade}, \citenamefont
  {Pimpinelli},\ and\ \citenamefont {Cadoret}}]{Aujol2001}%
  \BibitemOpen
  \bibfield  {author} {\bibinfo {author} {\bibfnamefont {E.}~\bibnamefont
  {Aujol}}, \bibinfo {author} {\bibfnamefont {A.}~\bibnamefont {Trassoudaine}},
  \bibinfo {author} {\bibfnamefont {L.}~\bibnamefont {Siozade}}, \bibinfo
  {author} {\bibfnamefont {A.}~\bibnamefont {Pimpinelli}}, \ and\ \bibinfo
  {author} {\bibfnamefont {R.}~\bibnamefont {Cadoret}},\ }\href@noop {}
  {\bibfield  {journal} {\bibinfo  {journal} {J. Cryst. Growth}\ }\textbf
  {\bibinfo {volume} {230}},\ \bibinfo {pages} {372} (\bibinfo {year}
  {2001})}\BibitemShut {NoStop}%
\bibitem [{\citenamefont {Chiang}\ \emph {et~al.}(1995)\citenamefont {Chiang},
  \citenamefont {Gates}, \citenamefont {Bensaoula},\ and\ \citenamefont
  {Schultz}}]{Chiang1995}%
  \BibitemOpen
  \bibfield  {author} {\bibinfo {author} {\bibfnamefont {C.-M.}\ \bibnamefont
  {Chiang}}, \bibinfo {author} {\bibfnamefont {S.~M.}\ \bibnamefont {Gates}},
  \bibinfo {author} {\bibfnamefont {A.}~\bibnamefont {Bensaoula}}, \ and\
  \bibinfo {author} {\bibfnamefont {J.~A.}\ \bibnamefont {Schultz}},\
  }\href@noop {} {\bibfield  {journal} {\bibinfo  {journal} {Chem. Phys.
  Lett.}\ }\textbf {\bibinfo {volume} {246}},\ \bibinfo {pages} {275} (\bibinfo
  {year} {1995})}\BibitemShut {NoStop}%
\bibitem [{\citenamefont {Shekhar}\ and\ \citenamefont
  {Jensen}(1997)}]{Shekhar1997}%
  \BibitemOpen
  \bibfield  {author} {\bibinfo {author} {\bibfnamefont {R.}~\bibnamefont
  {Shekhar}}\ and\ \bibinfo {author} {\bibfnamefont {K.~F.}\ \bibnamefont
  {Jensen}},\ }\href@noop {} {\bibfield  {journal} {\bibinfo  {journal} {Surf.
  Sci.}\ }\textbf {\bibinfo {volume} {381}},\ \bibinfo {pages} {L581} (\bibinfo
  {year} {1997})}\BibitemShut {NoStop}%
\bibitem [{\citenamefont {Bellitto}\ \emph
  {et~al.}(1999{\natexlab{a}})\citenamefont {Bellitto}, \citenamefont {Thoms},
  \citenamefont {Koleske}, \citenamefont {Wickenden},\ and\ \citenamefont
  {Henry}}]{bell99}%
  \BibitemOpen
  \bibfield  {author} {\bibinfo {author} {\bibfnamefont {V.~J.}\ \bibnamefont
  {Bellitto}}, \bibinfo {author} {\bibfnamefont {B.~D.}\ \bibnamefont {Thoms}},
  \bibinfo {author} {\bibfnamefont {D.~D.}\ \bibnamefont {Koleske}}, \bibinfo
  {author} {\bibfnamefont {A.~E.}\ \bibnamefont {Wickenden}}, \ and\ \bibinfo
  {author} {\bibfnamefont {R.~L.}\ \bibnamefont {Henry}},\ }\href@noop {}
  {\bibfield  {journal} {\bibinfo  {journal} {Surf. Sci.}\ }\textbf {\bibinfo
  {volume} {430}},\ \bibinfo {pages} {80} (\bibinfo {year}
  {1999}{\natexlab{a}})}\BibitemShut {NoStop}%
\bibitem [{\citenamefont {Bellitto}\ \emph
  {et~al.}(1999{\natexlab{b}})\citenamefont {Bellitto}, \citenamefont {Thoms},
  \citenamefont {Koleske}, \citenamefont {Wickenden},\ and\ \citenamefont
  {Henry}}]{bell99_b}%
  \BibitemOpen
  \bibfield  {author} {\bibinfo {author} {\bibfnamefont {V.~J.}\ \bibnamefont
  {Bellitto}}, \bibinfo {author} {\bibfnamefont {B.~D.}\ \bibnamefont {Thoms}},
  \bibinfo {author} {\bibfnamefont {D.~D.}\ \bibnamefont {Koleske}}, \bibinfo
  {author} {\bibfnamefont {A.~E.}\ \bibnamefont {Wickenden}}, \ and\ \bibinfo
  {author} {\bibfnamefont {R.~L.}\ \bibnamefont {Henry}},\ }\href@noop {}
  {\bibfield  {journal} {\bibinfo  {journal} {Phys. Rev. B}\ }\textbf {\bibinfo
  {volume} {60}},\ \bibinfo {pages} {4816} (\bibinfo {year}
  {1999}{\natexlab{b}})}\BibitemShut {NoStop}%
\bibitem [{\citenamefont {Sloboshanin}\ \emph {et~al.}(1999)\citenamefont
  {Sloboshanin}, \citenamefont {Tautz}, \citenamefont {Polyakov}, \citenamefont
  {Starke}, \citenamefont {Usikov}, \citenamefont {Ber},\ and\ \citenamefont
  {Schaefer}}]{slo99}%
  \BibitemOpen
  \bibfield  {author} {\bibinfo {author} {\bibfnamefont {S.}~\bibnamefont
  {Sloboshanin}}, \bibinfo {author} {\bibfnamefont {F.~S.}\ \bibnamefont
  {Tautz}}, \bibinfo {author} {\bibfnamefont {V.~M.}\ \bibnamefont {Polyakov}},
  \bibinfo {author} {\bibfnamefont {U.}~\bibnamefont {Starke}}, \bibinfo
  {author} {\bibfnamefont {A.~S.}\ \bibnamefont {Usikov}}, \bibinfo {author}
  {\bibfnamefont {B.~J.}\ \bibnamefont {Ber}}, \ and\ \bibinfo {author}
  {\bibfnamefont {J.~A.}\ \bibnamefont {Schaefer}},\ }\href@noop {} {\bibfield
  {journal} {\bibinfo  {journal} {Surf. Sci.}\ }\textbf {\bibinfo {volume}
  {427-428}},\ \bibinfo {pages} {250} (\bibinfo {year} {1999})}\BibitemShut
  {NoStop}%
\bibitem [{\citenamefont {Grabowski}\ \emph {et~al.}(2000)\citenamefont
  {Grabowski}, \citenamefont {Nienhaus},\ and\ \citenamefont
  {M{\"o}nch}}]{gra2000}%
  \BibitemOpen
  \bibfield  {author} {\bibinfo {author} {\bibfnamefont {S.~P.}\ \bibnamefont
  {Grabowski}}, \bibinfo {author} {\bibfnamefont {H.}~\bibnamefont {Nienhaus}},
  \ and\ \bibinfo {author} {\bibfnamefont {W.}~\bibnamefont {M{\"o}nch}},\
  }\href@noop {} {\bibfield  {journal} {\bibinfo  {journal} {Eur. Phys. J. B}\
  }\textbf {\bibinfo {volume} {16}},\ \bibinfo {pages} {3} (\bibinfo {year}
  {2000})}\BibitemShut {NoStop}%
\bibitem [{\citenamefont {Starke}\ \emph {et~al.}(2000)\citenamefont {Starke},
  \citenamefont {Sloboshanin}, \citenamefont {Tautz}, \citenamefont {Seubert},\
  and\ \citenamefont {Schaefer}}]{star00}%
  \BibitemOpen
  \bibfield  {author} {\bibinfo {author} {\bibfnamefont {U.}~\bibnamefont
  {Starke}}, \bibinfo {author} {\bibfnamefont {S.}~\bibnamefont {Sloboshanin}},
  \bibinfo {author} {\bibfnamefont {F.~S.}\ \bibnamefont {Tautz}}, \bibinfo
  {author} {\bibfnamefont {A.}~\bibnamefont {Seubert}}, \ and\ \bibinfo
  {author} {\bibfnamefont {J.A.}\ \bibnamefont {Schaefer}},\ }\href@noop {}
  {\bibfield  {journal} {\bibinfo  {journal} {Phys. Status Solidi A}\ }\textbf
  {\bibinfo {volume} {177}},\ \bibinfo {pages} {5} (\bibinfo {year}
  {2000})}\BibitemShut {NoStop}%
\bibitem [{\citenamefont {Pignedoli}\ \emph {et~al.}(2001)\citenamefont
  {Pignedoli}, \citenamefont {Di~Felice},\ and\ \citenamefont
  {Bertoni}}]{Pignedoli2001}%
  \BibitemOpen
  \bibfield  {author} {\bibinfo {author} {\bibfnamefont {C.~A.}\ \bibnamefont
  {Pignedoli}}, \bibinfo {author} {\bibfnamefont {R.}~\bibnamefont
  {Di~Felice}}, \ and\ \bibinfo {author} {\bibfnamefont {C.~M.}\ \bibnamefont
  {Bertoni}},\ }\href@noop {} {\bibfield  {journal} {\bibinfo  {journal} {Phys.
  Rev. B}\ }\textbf {\bibinfo {volume} {64}},\ \bibinfo {pages} {113301}
  (\bibinfo {year} {2001})}\BibitemShut {NoStop}%
\bibitem [{\citenamefont {de~Walle}\ and\ \citenamefont
  {Neugebauer}(2002)}]{Walle2002}%
  \BibitemOpen
  \bibfield  {author} {\bibinfo {author} {\bibfnamefont {C.~G.~Van}\
  \bibnamefont {de~Walle}}\ and\ \bibinfo {author} {\bibfnamefont
  {J.}~\bibnamefont {Neugebauer}},\ }\href@noop {} {\bibfield  {journal}
  {\bibinfo  {journal} {J. Vac. Sci. Technol. B}\ }\textbf {\bibinfo {volume}
  {20}},\ \bibinfo {pages} {1640} (\bibinfo {year} {2002})}\BibitemShut
  {NoStop}%
\bibitem [{\citenamefont {Northrup}\ and\ \citenamefont
  {Neugebauer}(2004)}]{Northrup2004}%
  \BibitemOpen
  \bibfield  {author} {\bibinfo {author} {\bibfnamefont {J.~E.}\ \bibnamefont
  {Northrup}}\ and\ \bibinfo {author} {\bibfnamefont {J.}~\bibnamefont
  {Neugebauer}},\ }\href@noop {} {\bibfield  {journal} {\bibinfo  {journal}
  {Appl. Phys. Lett.}\ }\textbf {\bibinfo {volume} {85}},\ \bibinfo {pages}
  {3429} (\bibinfo {year} {2004})}\BibitemShut {NoStop}%
\bibitem [{\citenamefont {Bermudez}(2004)}]{Bermudez2004}%
  \BibitemOpen
  \bibfield  {author} {\bibinfo {author} {\bibfnamefont {V.~M.}\ \bibnamefont
  {Bermudez}},\ }\href@noop {} {\bibfield  {journal} {\bibinfo  {journal}
  {Surf. Sci.}\ }\textbf {\bibinfo {volume} {565}},\ \bibinfo {pages} {89}
  (\bibinfo {year} {2004})}\BibitemShut {NoStop}%
\bibitem [{\citenamefont {Chen}\ \emph {et~al.}(2010)\citenamefont {Chen},
  \citenamefont {Sun},\ and\ \citenamefont {Hayashi}}]{Chen2010}%
  \BibitemOpen
  \bibfield  {author} {\bibinfo {author} {\bibfnamefont {P.-T.}\ \bibnamefont
  {Chen}}, \bibinfo {author} {\bibfnamefont {C.-L.}\ \bibnamefont {Sun}}, \
  and\ \bibinfo {author} {\bibfnamefont {M.}~\bibnamefont {Hayashi}},\
  }\href@noop {} {\bibfield  {journal} {\bibinfo  {journal} {J. Phys. Chem. C}\
  }\textbf {\bibinfo {volume} {114}},\ \bibinfo {pages} {18228} (\bibinfo
  {year} {2010})}\BibitemShut {NoStop}%
\bibitem [{\citenamefont {Kempisty}\ \emph {et~al.}(2011)\citenamefont
  {Kempisty}, \citenamefont {Strak},\ and\ \citenamefont
  {Krukowski}}]{Kempisty2011}%
  \BibitemOpen
  \bibfield  {author} {\bibinfo {author} {\bibfnamefont {P.}~\bibnamefont
  {Kempisty}}, \bibinfo {author} {\bibfnamefont {P.}~\bibnamefont {Strak}}, \
  and\ \bibinfo {author} {\bibfnamefont {S.}~\bibnamefont {Krukowski}},\
  }\href@noop {} {\bibfield  {journal} {\bibinfo  {journal} {Surf. Sci.}\
  }\textbf {\bibinfo {volume} {605}},\ \bibinfo {pages} {695} (\bibinfo {year}
  {2011})}\BibitemShut {NoStop}%
\bibitem [{\citenamefont {Kempisty}\ and\ \citenamefont
  {Krukowski}(2012)}]{Kempisty2012}%
  \BibitemOpen
  \bibfield  {author} {\bibinfo {author} {\bibfnamefont {P.}~\bibnamefont
  {Kempisty}}\ and\ \bibinfo {author} {\bibfnamefont {S.}~\bibnamefont
  {Krukowski}},\ }\href@noop {} {\bibfield  {journal} {\bibinfo  {journal} {J.
  Cryst. Growth}\ }\textbf {\bibinfo {volume} {358}},\ \bibinfo {pages} {64}
  (\bibinfo {year} {2012})}\BibitemShut {NoStop}%
\bibitem [{\citenamefont {Ptasinska}\ \emph {et~al.}(2015)\citenamefont
  {Ptasinska}, \citenamefont {Piechota},\ and\ \citenamefont
  {Krukowski}}]{Ptasinska2015}%
  \BibitemOpen
  \bibfield  {author} {\bibinfo {author} {\bibfnamefont {M.}~\bibnamefont
  {Ptasinska}}, \bibinfo {author} {\bibfnamefont {J.}~\bibnamefont {Piechota}},
  \ and\ \bibinfo {author} {\bibfnamefont {S.}~\bibnamefont {Krukowski}},\
  }\href@noop {} {\bibfield  {journal} {\bibinfo  {journal} {J. Phys. Chem. C}\
  }\textbf {\bibinfo {volume} {119}},\ \bibinfo {pages} {11563} (\bibinfo
  {year} {2015})}\BibitemShut {NoStop}%
\bibitem [{\citenamefont {Dreyer}\ \emph {et~al.}(2014)\citenamefont {Dreyer},
  \citenamefont {Janotti},\ and\ \citenamefont {Van~de Walle}}]{Dreyer2014}%
  \BibitemOpen
  \bibfield  {author} {\bibinfo {author} {\bibfnamefont {C.~E.}\ \bibnamefont
  {Dreyer}}, \bibinfo {author} {\bibfnamefont {A.}~\bibnamefont {Janotti}}, \
  and\ \bibinfo {author} {\bibfnamefont {C.~G.}\ \bibnamefont {Van~de Walle}},\
  }\href@noop {} {\bibfield  {journal} {\bibinfo  {journal} {Phys. Rev. B}\
  }\textbf {\bibinfo {volume} {89}},\ \bibinfo {pages} {081305} (\bibinfo
  {year} {2014})}\BibitemShut {NoStop}%
\bibitem [{\citenamefont {Northrup}\ and\ \citenamefont
  {Neugebauer}(1996)}]{Nor96-PhysRevB.53.R10477}%
  \BibitemOpen
  \bibfield  {author} {\bibinfo {author} {\bibfnamefont {J.~E.}\ \bibnamefont
  {Northrup}}\ and\ \bibinfo {author} {\bibfnamefont {J.}~\bibnamefont
  {Neugebauer}},\ }\href@noop {} {\bibfield  {journal} {\bibinfo  {journal}
  {Phys. Rev. B}\ }\textbf {\bibinfo {volume} {53}},\ \bibinfo {pages} {R10477}
  (\bibinfo {year} {1996})}\BibitemShut {NoStop}%
\bibitem [{\citenamefont {Segev}\ and\ \citenamefont {{Van de
  Walle}}(2007)}]{Segev2007L15}%
  \BibitemOpen
  \bibfield  {author} {\bibinfo {author} {\bibfnamefont {D.}~\bibnamefont
  {Segev}}\ and\ \bibinfo {author} {\bibfnamefont {C.~G.}\ \bibnamefont {{Van
  de Walle}}},\ }\href@noop {} {\bibfield  {journal} {\bibinfo  {journal}
  {Surf. Sci.}\ }\textbf {\bibinfo {volume} {601}},\ \bibinfo {pages} {L15}
  (\bibinfo {year} {2007})}\BibitemShut {NoStop}%
\bibitem [{\citenamefont {Segev}\ and\ \citenamefont {{Van de
  Walle}}(2006)}]{seg06}%
  \BibitemOpen
  \bibfield  {author} {\bibinfo {author} {\bibfnamefont {D.}~\bibnamefont
  {Segev}}\ and\ \bibinfo {author} {\bibfnamefont {C.~G.}\ \bibnamefont {{Van
  de Walle}}},\ }\href@noop {} {\bibfield  {journal} {\bibinfo  {journal}
  {Europhys. Lett.}\ }\textbf {\bibinfo {volume} {76}},\ \bibinfo {pages} {305}
  (\bibinfo {year} {2006})}\BibitemShut {NoStop}%
\bibitem [{\citenamefont {Bertelli}\ \emph {et~al.}(2009)\citenamefont
  {Bertelli}, \citenamefont {L\"optien}, \citenamefont {Wenderoth},
  \citenamefont {Rizzi}, \citenamefont {Ulbrich}, \citenamefont {Righi},
  \citenamefont {Ferretti}, \citenamefont {Martin-Samos}, \citenamefont
  {Bertoni},\ and\ \citenamefont {Catellani}}]{ber09}%
  \BibitemOpen
  \bibfield  {author} {\bibinfo {author} {\bibfnamefont {M.}~\bibnamefont
  {Bertelli}}, \bibinfo {author} {\bibfnamefont {P.}~\bibnamefont {L\"optien}},
  \bibinfo {author} {\bibfnamefont {M.}~\bibnamefont {Wenderoth}}, \bibinfo
  {author} {\bibfnamefont {A.}~\bibnamefont {Rizzi}}, \bibinfo {author}
  {\bibfnamefont {R.~G.}\ \bibnamefont {Ulbrich}}, \bibinfo {author}
  {\bibfnamefont {M.~C.}\ \bibnamefont {Righi}}, \bibinfo {author}
  {\bibfnamefont {A.}~\bibnamefont {Ferretti}}, \bibinfo {author}
  {\bibfnamefont {L.}~\bibnamefont {Martin-Samos}}, \bibinfo {author}
  {\bibfnamefont {C.~M.}\ \bibnamefont {Bertoni}}, \ and\ \bibinfo {author}
  {\bibfnamefont {A.}~\bibnamefont {Catellani}},\ }\href@noop {} {\bibfield
  {journal} {\bibinfo  {journal} {Phys. Rev. B}\ }\textbf {\bibinfo {volume}
  {80}},\ \bibinfo {pages} {115324} (\bibinfo {year} {2009})}\BibitemShut
  {NoStop}%
\bibitem [{\citenamefont {Lymperakis}\ \emph {et~al.}(2013)\citenamefont
  {Lymperakis}, \citenamefont {Weidlich}, \citenamefont {Eisele}, \citenamefont
  {Schnedler}, \citenamefont {Nys}, \citenamefont {Grandidier}, \citenamefont
  {Sti{\'e}venard}, \citenamefont {Dunin-Borkowski}, \citenamefont
  {Neugebauer},\ and\ \citenamefont {Ebert}}]{Lym_APL2013}%
  \BibitemOpen
  \bibfield  {author} {\bibinfo {author} {\bibfnamefont {L.}~\bibnamefont
  {Lymperakis}}, \bibinfo {author} {\bibfnamefont {P.~H.}\ \bibnamefont
  {Weidlich}}, \bibinfo {author} {\bibfnamefont {H.}~\bibnamefont {Eisele}},
  \bibinfo {author} {\bibfnamefont {M.}~\bibnamefont {Schnedler}}, \bibinfo
  {author} {\bibfnamefont {J.-P.}\ \bibnamefont {Nys}}, \bibinfo {author}
  {\bibfnamefont {B.}~\bibnamefont {Grandidier}}, \bibinfo {author}
  {\bibfnamefont {D.}~\bibnamefont {Sti{\'e}venard}}, \bibinfo {author}
  {\bibfnamefont {R.~E.}\ \bibnamefont {Dunin-Borkowski}}, \bibinfo {author}
  {\bibfnamefont {J.}~\bibnamefont {Neugebauer}}, \ and\ \bibinfo {author}
  {\bibfnamefont {P.}~\bibnamefont {Ebert}},\ }\href@noop {} {\bibfield
  {journal} {\bibinfo  {journal} {Appl. Phys. Lett.}\ }\textbf {\bibinfo
  {volume} {103}},\ \bibinfo {pages} {152101} (\bibinfo {year}
  {2013})}\BibitemShut {NoStop}%
\bibitem [{\citenamefont {Himmerlich}\ \emph {et~al.}(2014)\citenamefont
  {Himmerlich}, \citenamefont {Eisenhardt}, \citenamefont {Shokhovets},
  \citenamefont {Krischok}, \citenamefont {R{\"{a}}thel}, \citenamefont
  {Speiser}, \citenamefont {Neumann}, \citenamefont {Navarro-Quezada},\ and\
  \citenamefont {Esser}}]{him_APL2014}%
  \BibitemOpen
  \bibfield  {author} {\bibinfo {author} {\bibfnamefont {M.}~\bibnamefont
  {Himmerlich}}, \bibinfo {author} {\bibfnamefont {A.}~\bibnamefont
  {Eisenhardt}}, \bibinfo {author} {\bibfnamefont {S.}~\bibnamefont
  {Shokhovets}}, \bibinfo {author} {\bibfnamefont {S.}~\bibnamefont
  {Krischok}}, \bibinfo {author} {\bibfnamefont {J.}~\bibnamefont
  {R{\"{a}}thel}}, \bibinfo {author} {\bibfnamefont {E.}~\bibnamefont
  {Speiser}}, \bibinfo {author} {\bibfnamefont {M.~D.}\ \bibnamefont
  {Neumann}}, \bibinfo {author} {\bibfnamefont {A.}~\bibnamefont
  {Navarro-Quezada}}, \ and\ \bibinfo {author} {\bibfnamefont {N.}~\bibnamefont
  {Esser}},\ }\href@noop {} {\bibfield  {journal} {\bibinfo  {journal} {Appl.
  Phys. Lett.}\ }\textbf {\bibinfo {volume} {104}},\ \bibinfo {pages} {171602}
  (\bibinfo {year} {2014})}\BibitemShut {NoStop}%
\bibitem [{\citenamefont {Northrup}\ \emph {et~al.}(1997)\citenamefont
  {Northrup}, \citenamefont {Felice},\ and\ \citenamefont
  {Neugebauer}}]{nor97}%
  \BibitemOpen
  \bibfield  {author} {\bibinfo {author} {\bibfnamefont {J.~E.}\ \bibnamefont
  {Northrup}}, \bibinfo {author} {\bibfnamefont {R.~Di}\ \bibnamefont
  {Felice}}, \ and\ \bibinfo {author} {\bibfnamefont {J.}~\bibnamefont
  {Neugebauer}},\ }\href@noop {} {\bibfield  {journal} {\bibinfo  {journal}
  {Phys. Rev. B}\ }\textbf {\bibinfo {volume} {56}},\ \bibinfo {pages} {R4325}
  (\bibinfo {year} {1997})}\BibitemShut {NoStop}%
\bibitem [{\citenamefont {Akiyama}\ \emph {et~al.}(2011)\citenamefont
  {Akiyama}, \citenamefont {Yamashita}, \citenamefont {Nakamura},\ and\
  \citenamefont {Ito}}]{Akiyama2011}%
  \BibitemOpen
  \bibfield  {author} {\bibinfo {author} {\bibfnamefont {T.}~\bibnamefont
  {Akiyama}}, \bibinfo {author} {\bibfnamefont {T.}~\bibnamefont {Yamashita}},
  \bibinfo {author} {\bibfnamefont {K.}~\bibnamefont {Nakamura}}, \ and\
  \bibinfo {author} {\bibfnamefont {T.}~\bibnamefont {Ito}},\ }\href@noop {}
  {\bibfield  {journal} {\bibinfo  {journal} {J. Cryst. Growth}\ }\textbf
  {\bibinfo {volume} {318}},\ \bibinfo {pages} {79} (\bibinfo {year}
  {2011})}\BibitemShut {NoStop}%
\bibitem [{\citenamefont {Shen}\ \emph {et~al.}(2009)\citenamefont {Shen},
  \citenamefont {Allen}, \citenamefont {Hybertsen},\ and\ \citenamefont
  {Muckerman}}]{shen2009}%
  \BibitemOpen
  \bibfield  {author} {\bibinfo {author} {\bibfnamefont {X.}~\bibnamefont
  {Shen}}, \bibinfo {author} {\bibfnamefont {P.~B.}\ \bibnamefont {Allen}},
  \bibinfo {author} {\bibfnamefont {M.~S.}\ \bibnamefont {Hybertsen}}, \ and\
  \bibinfo {author} {\bibfnamefont {J.~T.}\ \bibnamefont {Muckerman}},\
  }\href@noop {} {\bibfield  {journal} {\bibinfo  {journal} {J. Phys. Chem. C}\
  }\textbf {\bibinfo {volume} {113}},\ \bibinfo {pages} {3365} (\bibinfo {year}
  {2009})}\BibitemShut {NoStop}%
\bibitem [{\citenamefont {Paskova}\ \emph {et~al.}(2009)\citenamefont
  {Paskova}, \citenamefont {Preble}, \citenamefont {Hanser}, \citenamefont
  {Evans}, \citenamefont {Kr\"{o}ger}, \citenamefont {Paskov}, \citenamefont
  {Cheng}, \citenamefont {Park}, \citenamefont {Grenko},\ and\ \citenamefont
  {Johnson}}]{pas09}%
  \BibitemOpen
  \bibfield  {author} {\bibinfo {author} {\bibfnamefont {T.}~\bibnamefont
  {Paskova}}, \bibinfo {author} {\bibfnamefont {E.~A.}\ \bibnamefont {Preble}},
  \bibinfo {author} {\bibfnamefont {A.~D.}\ \bibnamefont {Hanser}}, \bibinfo
  {author} {\bibfnamefont {K.~R.}\ \bibnamefont {Evans}}, \bibinfo {author}
  {\bibfnamefont {R.}~\bibnamefont {Kr\"{o}ger}}, \bibinfo {author}
  {\bibfnamefont {P.~P.}\ \bibnamefont {Paskov}}, \bibinfo {author}
  {\bibfnamefont {A.~J.}\ \bibnamefont {Cheng}}, \bibinfo {author}
  {\bibfnamefont {M.}~\bibnamefont {Park}}, \bibinfo {author} {\bibfnamefont
  {J.~A.}\ \bibnamefont {Grenko}}, \ and\ \bibinfo {author} {\bibfnamefont
  {M.~A.~L.}\ \bibnamefont {Johnson}},\ }\href@noop {} {\bibfield  {journal}
  {\bibinfo  {journal} {Phys. Status Solidi C}\ }\textbf {\bibinfo {volume}
  {6}},\ \bibinfo {pages} {S344} (\bibinfo {year} {2009})}\BibitemShut
  {NoStop}%
\bibitem [{\citenamefont {Himmerlich}\ \emph {et~al.}(2007)\citenamefont
  {Himmerlich}, \citenamefont {Krischok}, \citenamefont {Lebedev},
  \citenamefont {Ambacher},\ and\ \citenamefont {Schaefer}}]{him07}%
  \BibitemOpen
  \bibfield  {author} {\bibinfo {author} {\bibfnamefont {M.}~\bibnamefont
  {Himmerlich}}, \bibinfo {author} {\bibfnamefont {S.}~\bibnamefont
  {Krischok}}, \bibinfo {author} {\bibfnamefont {V.}~\bibnamefont {Lebedev}},
  \bibinfo {author} {\bibfnamefont {O.}~\bibnamefont {Ambacher}}, \ and\
  \bibinfo {author} {\bibfnamefont {J.~A.}\ \bibnamefont {Schaefer}},\
  }\href@noop {} {\bibfield  {journal} {\bibinfo  {journal} {J. Cryst. Growth}\
  }\textbf {\bibinfo {volume} {306}},\ \bibinfo {pages} {6} (\bibinfo {year}
  {2007})}\BibitemShut {NoStop}%
\bibitem [{\citenamefont {Ibach}(1993)}]{IBACH1993819}%
  \BibitemOpen
  \bibfield  {author} {\bibinfo {author} {\bibfnamefont {H.}~\bibnamefont
  {Ibach}},\ }\bibfield  {title} {\enquote {\bibinfo {title} {Electron energy
  loss spectroscopy with resolution below 1 mev},}\ }\href@noop {} {\bibfield
  {journal} {\bibinfo  {journal} {J. Electron Spectrosc.}\ }\textbf {\bibinfo
  {volume} {64}},\ \bibinfo {pages} {819} (\bibinfo {year} {1993})}\BibitemShut
  {NoStop}%
\bibitem [{\citenamefont {Kresse}\ and\ \citenamefont
  {Hafner}(1993)}]{Kresse93}%
  \BibitemOpen
  \bibfield  {author} {\bibinfo {author} {\bibfnamefont {G.}~\bibnamefont
  {Kresse}}\ and\ \bibinfo {author} {\bibfnamefont {J.}~\bibnamefont
  {Hafner}},\ }\href@noop {} {\bibfield  {journal} {\bibinfo  {journal} {Phys.
  Rev. B}\ }\textbf {\bibinfo {volume} {47}},\ \bibinfo {pages} {558} (\bibinfo
  {year} {1993})}\BibitemShut {NoStop}%
\bibitem [{\citenamefont {Kresse}\ and\ \citenamefont
  {Furthm\"{u}ller}(1996)}]{Kresse96}%
  \BibitemOpen
  \bibfield  {author} {\bibinfo {author} {\bibfnamefont {G.}~\bibnamefont
  {Kresse}}\ and\ \bibinfo {author} {\bibfnamefont {J.}~\bibnamefont
  {Furthm\"{u}ller}},\ }\href@noop {} {\bibfield  {journal} {\bibinfo
  {journal} {Phys. Rev. B}\ }\textbf {\bibinfo {volume} {54}},\ \bibinfo
  {pages} {11169} (\bibinfo {year} {1996})}\BibitemShut {NoStop}%
\bibitem [{\citenamefont {Duff}\ \emph {et~al.}(2014)\citenamefont {Duff},
  \citenamefont {Lymperakis},\ and\ \citenamefont {Neugebauer}}]{Duff2014}%
  \BibitemOpen
  \bibfield  {author} {\bibinfo {author} {\bibfnamefont {A.~I.}\ \bibnamefont
  {Duff}}, \bibinfo {author} {\bibfnamefont {L.}~\bibnamefont {Lymperakis}}, \
  and\ \bibinfo {author} {\bibfnamefont {J.}~\bibnamefont {Neugebauer}},\
  }\href@noop {} {\bibfield  {journal} {\bibinfo  {journal} {Phys. Rev. B}\
  }\textbf {\bibinfo {volume} {89}},\ \bibinfo {pages} {085307} (\bibinfo
  {year} {2014})}\BibitemShut {NoStop}%
\bibitem [{\citenamefont {Vineyard}(1957)}]{Vineyard1957}%
  \BibitemOpen
  \bibfield  {author} {\bibinfo {author} {\bibfnamefont {G.~H.}\ \bibnamefont
  {Vineyard}},\ }\href@noop {} {\bibfield  {journal} {\bibinfo  {journal} {J.
  Phys. Chem. Solids}\ }\textbf {\bibinfo {volume} {3}},\ \bibinfo {pages}
  {121} (\bibinfo {year} {1957})}\BibitemShut {NoStop}%
\bibitem [{\citenamefont {Henkelman}\ \emph {et~al.}(2000)\citenamefont
  {Henkelman}, \citenamefont {Uberuaga},\ and\ \citenamefont
  {J\'{o}nsson}}]{NEB}%
  \BibitemOpen
  \bibfield  {author} {\bibinfo {author} {\bibfnamefont {G.}~\bibnamefont
  {Henkelman}}, \bibinfo {author} {\bibfnamefont {B.~P.}\ \bibnamefont
  {Uberuaga}}, \ and\ \bibinfo {author} {\bibfnamefont {H.}~\bibnamefont
  {J\'{o}nsson}},\ }\href@noop {} {\bibfield  {journal} {\bibinfo  {journal}
  {J. Chem. Phys.}\ }\textbf {\bibinfo {volume} {113}},\ \bibinfo {pages}
  {9901} (\bibinfo {year} {2000})}\BibitemShut {NoStop}%
\bibitem [{\citenamefont {Heyd}\ \emph {et~al.}(2003)\citenamefont {Heyd},
  \citenamefont {Scuseria},\ and\ \citenamefont {Ernzerhof}}]{HSE}%
  \BibitemOpen
  \bibfield  {author} {\bibinfo {author} {\bibfnamefont {J.}~\bibnamefont
  {Heyd}}, \bibinfo {author} {\bibfnamefont {G.~E.}\ \bibnamefont {Scuseria}},
  \ and\ \bibinfo {author} {\bibfnamefont {M.}~\bibnamefont {Ernzerhof}},\
  }\href@noop {} {\bibfield  {journal} {\bibinfo  {journal} {J. Chem. Phys.}\
  }\textbf {\bibinfo {volume} {118}},\ \bibinfo {pages} {8207} (\bibinfo {year}
  {2003})}\BibitemShut {NoStop}%
\bibitem [{\citenamefont {Cui}\ \emph {et~al.}(2015)\citenamefont {Cui},
  \citenamefont {Lee}, \citenamefont {Freysoldt},\ and\ \citenamefont
  {Neugebauer}}]{RN2097}%
  \BibitemOpen
  \bibfield  {author} {\bibinfo {author} {\bibfnamefont {Y.}~\bibnamefont
  {Cui}}, \bibinfo {author} {\bibfnamefont {S.}~\bibnamefont {Lee}}, \bibinfo
  {author} {\bibfnamefont {C.}~\bibnamefont {Freysoldt}}, \ and\ \bibinfo
  {author} {\bibfnamefont {J.}~\bibnamefont {Neugebauer}},\ }\href
  {http://link.aps.org/doi/10.1103/PhysRevB.92.085204} {\bibfield  {journal}
  {\bibinfo  {journal} {Phys. Rev. B}\ }\textbf {\bibinfo {volume} {92}},\
  \bibinfo {pages} {085204} (\bibinfo {year} {2015})}\BibitemShut {NoStop}%
\bibitem [{\citenamefont {Lambin}\ \emph {et~al.}(1985)\citenamefont {Lambin},
  \citenamefont {Vigneron},\ and\ \citenamefont {Lucas}}]{lam85}%
  \BibitemOpen
  \bibfield  {author} {\bibinfo {author} {\bibfnamefont {P.}~\bibnamefont
  {Lambin}}, \bibinfo {author} {\bibfnamefont {J.~P.}\ \bibnamefont
  {Vigneron}}, \ and\ \bibinfo {author} {\bibfnamefont {A.~A.}\ \bibnamefont
  {Lucas}},\ }\href@noop {} {\bibfield  {journal} {\bibinfo  {journal} {Phys.
  Rev. B}\ }\textbf {\bibinfo {volume} {32}},\ \bibinfo {pages} {8203}
  (\bibinfo {year} {1985})}\BibitemShut {NoStop}%
\bibitem [{\citenamefont {Lambin}\ \emph {et~al.}(1990)\citenamefont {Lambin},
  \citenamefont {Vigneron},\ and\ \citenamefont {Lucas}}]{lam90}%
  \BibitemOpen
  \bibfield  {author} {\bibinfo {author} {\bibfnamefont {P.}~\bibnamefont
  {Lambin}}, \bibinfo {author} {\bibfnamefont {J.~P.}\ \bibnamefont
  {Vigneron}}, \ and\ \bibinfo {author} {\bibfnamefont {A.~A.}\ \bibnamefont
  {Lucas}},\ }\href@noop {} {\bibfield  {journal} {\bibinfo  {journal} {Comput.
  Phys. Commun.}\ }\textbf {\bibinfo {volume} {60}},\ \bibinfo {pages} {351}
  (\bibinfo {year} {1990})}\BibitemShut {NoStop}%
\bibitem [{\citenamefont {Davydov}\ \emph {et~al.}(1998)\citenamefont
  {Davydov}, \citenamefont {Kitaev}, \citenamefont {Goncharuk}, \citenamefont
  {Smirnov}, \citenamefont {Graul}, \citenamefont {Semchinova}, \citenamefont
  {Uffmann}, \citenamefont {Smirnov}, \citenamefont {Mirgorodsky},\ and\
  \citenamefont {Evarestov}}]{PhysRevB.58.12899}%
  \BibitemOpen
  \bibfield  {author} {\bibinfo {author} {\bibfnamefont {V.~Y.}\ \bibnamefont
  {Davydov}}, \bibinfo {author} {\bibfnamefont {Y.~E.}\ \bibnamefont {Kitaev}},
  \bibinfo {author} {\bibfnamefont {I.~N.}\ \bibnamefont {Goncharuk}}, \bibinfo
  {author} {\bibfnamefont {A.~N.}\ \bibnamefont {Smirnov}}, \bibinfo {author}
  {\bibfnamefont {J.}~\bibnamefont {Graul}}, \bibinfo {author} {\bibfnamefont
  {O.}~\bibnamefont {Semchinova}}, \bibinfo {author} {\bibfnamefont
  {D.}~\bibnamefont {Uffmann}}, \bibinfo {author} {\bibfnamefont {M.~B.}\
  \bibnamefont {Smirnov}}, \bibinfo {author} {\bibfnamefont {A.~P.}\
  \bibnamefont {Mirgorodsky}}, \ and\ \bibinfo {author} {\bibfnamefont {R.~A.}\
  \bibnamefont {Evarestov}},\ }\href@noop {} {\bibfield  {journal} {\bibinfo
  {journal} {Phys. Rev. B}\ }\textbf {\bibinfo {volume} {58}},\ \bibinfo
  {pages} {12899} (\bibinfo {year} {1998})}\BibitemShut {NoStop}%
\bibitem [{\citenamefont {Grabowski}\ \emph {et~al.}(1996)\citenamefont
  {Grabowski}, \citenamefont {Nienhaus},\ and\ \citenamefont
  {M{\"o}nch}}]{gra96}%
  \BibitemOpen
  \bibfield  {author} {\bibinfo {author} {\bibfnamefont {S.~P.}\ \bibnamefont
  {Grabowski}}, \bibinfo {author} {\bibfnamefont {H.}~\bibnamefont {Nienhaus}},
  \ and\ \bibinfo {author} {\bibfnamefont {W.}~\bibnamefont {M{\"o}nch}},\
  }\href@noop {} {\bibfield  {journal} {\bibinfo  {journal} {Surf. Sci.}\
  }\textbf {\bibinfo {volume} {352-354}},\ \bibinfo {pages} {310} (\bibinfo
  {year} {1996})}\BibitemShut {NoStop}%
\bibitem [{\citenamefont {Landmann}\ \emph {et~al.}(2015)\citenamefont
  {Landmann}, \citenamefont {Rauls}, \citenamefont {Schmidt}, \citenamefont
  {Neumann}, \citenamefont {Speiser},\ and\ \citenamefont
  {Esser}}]{PhysRevB.91.035302}%
  \BibitemOpen
  \bibfield  {author} {\bibinfo {author} {\bibfnamefont {M.}~\bibnamefont
  {Landmann}}, \bibinfo {author} {\bibfnamefont {E.}~\bibnamefont {Rauls}},
  \bibinfo {author} {\bibfnamefont {W.~G.}\ \bibnamefont {Schmidt}}, \bibinfo
  {author} {\bibfnamefont {M.~D.}\ \bibnamefont {Neumann}}, \bibinfo {author}
  {\bibfnamefont {E.}~\bibnamefont {Speiser}}, \ and\ \bibinfo {author}
  {\bibfnamefont {N.}~\bibnamefont {Esser}},\ }\href {\doibase
  10.1103/PhysRevB.91.035302} {\bibfield  {journal} {\bibinfo  {journal} {Phys.
  Rev. B}\ }\textbf {\bibinfo {volume} {91}},\ \bibinfo {pages} {035302}
  (\bibinfo {year} {2015})}\BibitemShut {NoStop}%
\bibitem [{\citenamefont {{Van de Walle}}\ and\ \citenamefont
  {Segev}(2007)}]{wal07}%
  \BibitemOpen
  \bibfield  {author} {\bibinfo {author} {\bibfnamefont {C.~G.}\ \bibnamefont
  {{Van de Walle}}}\ and\ \bibinfo {author} {\bibfnamefont {D.}~\bibnamefont
  {Segev}},\ }\href@noop {} {\bibfield  {journal} {\bibinfo  {journal} {J.
  Appl. Phys.}\ }\textbf {\bibinfo {volume} {101}},\ \bibinfo {pages} {081704}
  (\bibinfo {year} {2007})}\BibitemShut {NoStop}%
\bibitem [{\citenamefont {Hertz}(1882)}]{ANDP:ANDP18822531002}%
  \BibitemOpen
  \bibfield  {author} {\bibinfo {author} {\bibfnamefont {H.}~\bibnamefont
  {Hertz}},\ }\href {\doibase 10.1002/andp.18822531002} {\bibfield  {journal}
  {\bibinfo  {journal} {Ann. Phys.}\ }\textbf {\bibinfo {volume} {253}},\
  \bibinfo {pages} {177} (\bibinfo {year} {1882})}\BibitemShut {NoStop}%
\bibitem [{\citenamefont {Knudsen}(1915)}]{ANDP:ANDP19153521306}%
  \BibitemOpen
  \bibfield  {author} {\bibinfo {author} {\bibfnamefont {M.}~\bibnamefont
  {Knudsen}},\ }\href {\doibase 10.1002/andp.19153521306} {\bibfield  {journal}
  {\bibinfo  {journal} {Ann. Phys.}\ }\textbf {\bibinfo {volume} {352}},\
  \bibinfo {pages} {697} (\bibinfo {year} {1915})}\BibitemShut {NoStop}%
\bibitem [{\citenamefont {Winkler}\ and\ \citenamefont
  {Rendulic}(1992)}]{WinklerRendulic}%
  \BibitemOpen
  \bibfield  {author} {\bibinfo {author} {\bibfnamefont {A.}~\bibnamefont
  {Winkler}}\ and\ \bibinfo {author} {\bibfnamefont {K.~D.}\ \bibnamefont
  {Rendulic}},\ }\href@noop {} {\bibfield  {journal} {\bibinfo  {journal} {Int.
  Rev. Phys. Chem.}\ }\textbf {\bibinfo {volume} {11}},\ \bibinfo {pages} {101}
  (\bibinfo {year} {1992})}\BibitemShut {NoStop}%
\bibitem [{\citenamefont {Steinr\"uck}\ \emph {et~al.}(1985)\citenamefont
  {Steinr\"uck}, \citenamefont {Luger}, \citenamefont {Winkler},\ and\
  \citenamefont {Rendulic}}]{PhysRevB.32.5032}%
  \BibitemOpen
  \bibfield  {author} {\bibinfo {author} {\bibfnamefont {H.~P.}\ \bibnamefont
  {Steinr\"uck}}, \bibinfo {author} {\bibfnamefont {M.}~\bibnamefont {Luger}},
  \bibinfo {author} {\bibfnamefont {A.}~\bibnamefont {Winkler}}, \ and\
  \bibinfo {author} {\bibfnamefont {K.~D.}\ \bibnamefont {Rendulic}},\ }\href
  {\doibase 10.1103/PhysRevB.32.5032} {\bibfield  {journal} {\bibinfo
  {journal} {Phys. Rev. B}\ }\textbf {\bibinfo {volume} {32}},\ \bibinfo
  {pages} {5032} (\bibinfo {year} {1985})}\BibitemShut {NoStop}%
\end{thebibliography}
\end{document}